\documentclass[pre,showkeys,11pt]{revtex4}

\usepackage{graphicx}
\usepackage{dcolumn}
\usepackage{bm}
\usepackage{amsmath}

\DeclareSymbolFont{AMSb}{U}{msb}{m}{n}
\DeclareMathSymbol{\N}{\mathbin}{AMSb}{"4E}
\DeclareMathSymbol{\Z}{\mathbin}{AMSb}{"5A}
\DeclareMathSymbol{\R}{\mathbin}{AMSb}{"52}
\DeclareMathSymbol{\Q}{\mathbin}{AMSb}{"51}
\DeclareMathSymbol{\I}{\mathbin}{AMSb}{"49}
\DeclareMathSymbol{\C}{\mathbin}{AMSb}{"43}

\newcommand{\inv}{^{\raise.15ex\hbox{${\scriptscriptstyle -}$}\kern-.05em 1}}
\def\pnss{j}
\def\kn{m} 
\def\moment{\tau}
\def\pnQ{Q}\def\pndis{_{\mathrm{discrete}}}
\def\ex#1{e^{#1}}
\def\vt{\vec{t}}\def\vti{\vec{t}_{\rm i}}\def\vtf{\vec{t}_{\rm f}}
\def\vu{\vec{u}}\def\vum#1{\vu_{#1}}
\def\vv{\vec{v}}\def\vvm#1{\vv_{#1}}
\def\vx{\vec{x}}
\def\vk{\vec{k}}
\def\eref#1{eqn~\ref{#1}}
\def\Eref#1{Eqn~\ref{#1}}
\def\sref#1{Sect.~\ref{#1}}\def\srefs#1{Sects.~\ref{#1}}
\def\fref#1{fig~\ref{#1}}\def\Fref#1{Fig~\ref{#1}}\def\frefs#1{figs~\ref{#1}}

\def\kbt{k_{\rm B}T}
\newcommand{\pnlabel}[1]{\label{#1}}
\def\xst{\xi^*}\def\xk{\xi_{\rm kink}}
\newcommand{\pnmeas}[1]{[d\vt\,(s)]_{#1}\,}
\newcommand{\pn}[2]{#1}

\begin{document}

\preprint{APS/123-QED}

\title{Exact theory of kinkable elastic polymers}

\author{Paul A. Wiggins}
\email{pwiggins@caltech.edu}
\altaffiliation{Division of Physics, Mathematics, \& Astronomy, California Institute of Technology}
 \homepage{http://www.rpgroup.caltech.edu/~wiggins}

\author{Rob Phillips}
\email{phillips@aero.caltech.edu}
\altaffiliation{Division of Engineering and Applied Science, California Institute of Technology.}

\author{Philip C. Nelson}
\email{nelson@physics.upenn.edu} \altaffiliation{Department of
Physics and Astronomy, University of Pennsylvania.}

\date{August 31st, 2004}

\begin{abstract}
    \noindent
The importance of nonlinearities in material constitutive relations has
long been appreciated in the continuum mechanics of macroscopic
rods
. Although the moment (torque) response to bending is almost
universally linear for small deflection angles, many rod systems
exhibit a high-curvature softening. The signature behavior of
these rod systems is a kinking transition in which the bending is
localized. Recent DNA cyclization experiments by Cloutier and
Widom
 have offered evidence that the
linear-elastic bending theory fails to describe the high-curvature
mechanics of DNA. Motivated by this recent experimental work, we
develop a simple and exact theory of the statistical mechanics of
linear-elastic polymer chains that can undergo a kinking
transition. We characterize the kinking behavior with a single
parameter and show that the resulting theory reproduces both the
low-curvature linear-elastic behavior which is already well described by
the Wormlike Chain model, as well as the high-curvature softening observed in recent cyclization experiments.
\end{abstract}


\maketitle

\section{Introduction}
The behavior of many semiflexible polymers is
captured by the Wormlike Chain model\cite{KP,SY}. \pn{This model amounts}{Although this
theory was originally proposed as a purely entropic theory of
freely rotating rigid bonds with small fixed bond angles\cite{KP},
it is equivalent} to the statistical mechanics of linearly-elastic
rods\cite{YamakawaBook} where the elastic energy is
microscopically a combination of both energetic and entropic
contributions\cite{PhilBook}. The mechanics of DNA, a polymer of
particular biological interest, has been studied extensively
experimentally and theoretically and its mechanical properties
have been very well approximated by the Wormlike Chain model
(WLC)\cite{Bustamante2000} and its successors such as the Helical Wormlike
Chain model\cite{SY}. For example, accurate force-extension
experiments have shown that DNA is surprisingly well described by
WLC\cite{Bustamante2000,Bouchiat,PhilBook}, at least until the
effects of DNA stretching become important at tensions of order 50
pN.

Despite the success of the WLC in describing DNA
mechanics, recent DNA cyclization experiments by Cloutier and
Widom\cite{CW} have shown a dramatic departure from theoretical
predictions for highly-curved DNA. These experiments suggest that
the effective bending energy of small, cyclized sequences of DNA is
significantly smaller than predicted by existing theoretical
models based upon linear-elastic constitutive relations, in which
the bending energy is quadratic in curvature. Similar anomalies
have been revealed in transcriptional regulation where DNA looping
by regulatory proteins remains active down to $60$ basepair (bp) separations
between the binding
sites\cite{bell88a,Rippe1995,Muller1,Muller2,Rippe2001,Zhang}.

From a continuum-mechanics perspective, this failure of the model at high-curvature is hardly surprising; the
importance of material nonlinearities has been appreciated for many years. In fact,
anyone who has ever tried to bend a drinking straw has observed that the straw will at first distribute
the bending,
as predicted by the linear theory, but as the curvature increases, the straw will eventually kink,
localizing the bending. This kinking behavior is the signature of
nonlinear constitutive softening at high curvature.
Nonlinearities are certainly important in microscopic physical systems, such as polymers,
because the effective bending free energy, a combination of interaction potentials and entropic
effects, is only approximately harmonic. The possibility of kinking
in DNA was realized long ago by Crick and Klug, who proposed a specific
atomistic structure for the kink state \cite{cric75a}. Many authors
have since found kinked states of DNA in protein--DNA complexes \pn{(see
for example \cite{dick98a}),}{} but
less attention has been given to \textit{spontaneous} kinking of free
DNA in solution, even though Crick and Klug pointed out this
possibility.

Our goal in this paper is to develop a simple, generic extension
of the WLC model, introducing only one additional parameter: the
average number of kinks per unit length for the unconstrained
chain. The ``kinks'' are taken to be freely-bending hinge elements
in the chain. This model is an extension of the well known
Wormlike Chain (WLC); we refer to it as the Kinkable Wormlike
Chain (KWLC). Although our model is not a detailed microscopic
picture for DNA, it does capture the key consequences of any more
detailed picture of kink formation. As such, it serves as a useful
coarse-grained model to describe high-curvature phenomena in many
stiff biopolymers, not just DNA \cite{Muroga}. \pn{Our main
results are summarized in \frefs{force.ext}, \ref{philfig},
\ref{widomdata}, and \ref{pndist}.}{}

The KWLC is the simplest example of a class of theories that have been
proposed and studied by Storm and Nelson\cite{Astor03b} and more
recently by Levine\cite{Levine}. It is simple enough that many
results are exact or nearly so. The method by which we obtain our
exact results is analogous to the Dyson expansion for time-dependent quantum
perturbation theory. For the KWLC, the perturbation
series can be re-summed exactly.

For small values of our kinking parameter the KWLC model predicts
nearly identical behavior to the WLC---except when the rod is constrained
to be highly curved. Such
constraints induce kinking, even when the kinking parameter is small.
We will show in detail how the energy relief caused by this
alternative bending conformation can account for the
observed anomalously high cyclization rate of short loops of
DNA\cite{CW} and anomalously high levels of gene
expression\cite{Muller1,Muller2}. Further discussion of the
applications of KWLC to
DNA, will appear elsewhere\cite{wiggins2}; the present paper focuses
on the mathematical details of the theory. Yan and Marko, and
Vologodskii, have independently obtained results related to ours
\cite{voloX,yan04a}. Also, Sucato et al.\ have performed Monte Carlo simulations of kinkable chains to obtain information about their
structural and thermodynamic properties \cite{suca04a}.

The outline of the paper is as follows: in section \ref{model}, we
introduce the KWLC model in a discrete form. In section
\ref{partitionfunction}, we compute the unconstrained partition
function for the theory and show that there is a sensible
continuum limit. In section
\ref{tanpart}, we give an exact computation of the tangent partition function
of the continuum theory as well the
moment-bend constitutive relation and the kink number for bent
polymer chains. We show that kinking causes an exact
renormalization of the tangent persistence length and we write
exact expressions for the average squared end distance and the
radius of gyration. In section \ref{tssp}, we exactly compute the
Fourier-Laplace transform of the spatial
propagator and discuss various limits of these results. We also
compute the exact force-extension relation and the structure
factor for KWLC. In section \ref{Jfactor}, we compute the KWLC
correction to the Jacobson-Stockmayer $J$ factor and the partition
function for cyclized chains. We show that the topological
constraint of cyclization induces kinking and we compute the kink
number distribution explicitly. In section \ref{discussion}, we
discuss the limitations of KWLC. In the appendix, we present a
summary of the Faltung Theorem which is required for computations
and develop the small and large contour length limits of the KWLC
$J$ factor.

\section{Kinkable Wormlike Chain Model}
\pnlabel{model} Although the Wormlike Chain model was originally
proposed to describe a purely entropic chain without a bending
energy\cite{KP}, it is often interpreted as the statistical
mechanics of rods with bending energies quadratic in
curvature\cite{YamakawaBook,LL}. From a mechanical perspective,
the success of the WLC model is not surprising since the small
amplitude bending of rods universally induces a linear moment
response. For WLC, the bending energy for a polymer in
configuration $\Gamma$ is
\begin{equation}
E_{\Gamma} = \int_{0}^L ds
\frac{\xi}{2}\left(\frac{d\vec{t}}{ds}\,\right)^2
\end{equation}
where $\vec{t}(s)$ is the unit tangent at arc length $s$, $L$ is the
contour length, and $\xi$ is the bending modulus. Throughout this
paper we will express energies in units of the room-temperature
thermal energy $\kbt=4.1\times10^{-21}\,$J. For WLC it
is well known that the bending modulus and persistence length (the
length scale over which tangent are thermally correlated) are
equal in these units \cite{PhilBook}.

It is most intuitive to define our new model in
terms of the discretized definition of WLC. Accordingly, we divide a
chain of arc length $L$ into $L/\ell$ segments of length $\ell$. There are then
$N=(L/\ell)-1$ interior vertices, plus two endpoints (\fref{disc.cont}a).
Next we replace the arc length derivative with the finite
difference over the segment length $\ell$, replace the integral with a
sum, and introduce the spring constant $\kappa\equiv
\xi/\ell$. The resulting energy is
\begin{equation}
E_{\Gamma} = \sum_{i=1}^{N} \kappa\left(1-\vec{t}_i\cdot\vec{t}_{i-1}\right),
\end{equation}
where $\vec{t}_i$ is the vector joining vertices $i$ and $i+1$.
\begin{figure}
\includegraphics{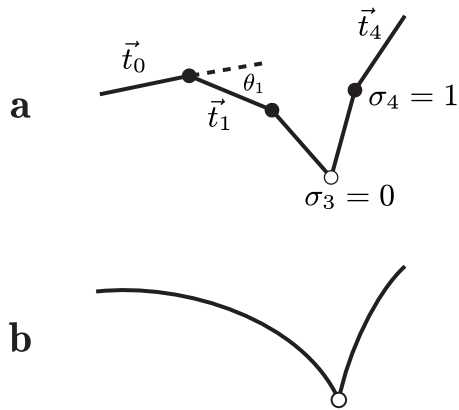}
\caption{\pnlabel{disc.cont}{\it a: } The discretized KWLC is a
chain of wormlike and kink-like vertices. In this illustration
$N=4$; thus there are four vertices, of which one is kink-like.
When a vertex $i$ is wormlike ($\sigma_i=1$), the energy is given
by the normal wormlike chain energy; if it is kink like
($\sigma_i=0$), the energy is $\epsilon$, independent of
$\theta_i$. {\it b: } The continuum version of this theory.
Although the number of vertices is now infinite, the continuum
limit maintains a finite average kink density.}
\end{figure}

We introduce a similar discretized
energy for the Kinkable Wormlike Chain model (KWLC).  In addition to the bending angle,
there is now a degree of freedom at each vertex describing whether the vertex is kink-like
or wormlike. To describe this degree of freedom, we introduce a state variable, $\sigma_i$ at each vertex.
When $\sigma_i=1$, we say that the vertex is
wormlike and the energy is given by the discrete WLC energy at that vertex. When $\sigma_i=0$, the vertex is kink
like and the energy is independent of the bend angle  at that vertex,
but there is an energy penalty $\epsilon$ to realize the kink state. This model is depicted schematically
in \fref{disc.cont}. The energy for the model we have just described can be concisely written as
\begin{equation}
E^*_{\Gamma} = \sum_{i=1}^{N}
\left[\kappa\left(1-\vec{t}_i\cdot\vec{t}_{i-1}\right)\sigma_i + \epsilon
(1-\sigma_i)\right]\ ,
\end{equation}
where the $*$ denotes that this is the energy of the KWLC theory
and $\epsilon$ is the energetic cost of introducing a kink in the
chain. Note that in general we denote KWLC results or equations with a $*$.
We will recover the WLC results when we take the kinking energy $\epsilon$ to
infinity. While Storm and Nelson\cite{Astor03a} and
others\cite{mark98a,Ahsan,CV,RouzinaBloomfield1,Astor03a,Levine}
have considered more general theories where the kink energy is not
assumed to be independent of the kink angle,  much of the
important physics can already be studied in the simpler KWLC
theory. Moreover, this theory has the significant advantage of
being analytically exact to a much greater extent than more
general theories; it applies in the limit where the kinks are only
weakly elastic compared to the elastic rod.

\section{Partition functions}\pnlabel{partitionfunction}
For a summary of notation used in this article, see
Appendix~\ref{notation}.

We have defined the KWLC model in terms of a discrete set of degrees
of freedom.   In the next section, however, we shall wish to take
advantage of the continuum WLC machinery. To this end, this section
formulates the continuum limit of
the KWLC model. Beyond the computational advantage, there is
also an additional reason to go to the continuum limit. \Fref{disc.cont} describes the kinking with two parameters, a
density $\ell\inv$ of kinkable sites  and the kink
energy $\epsilon$. We wish to describe the kinking in terms of a \textit{single}
parameter, to be called $\zeta$ (see \eref{zeta}). $\zeta$ essentially
sets the average number of kinks per contour length for a long,
unstressed chain.
In the continuum limit of WLC, we take $\ell\to0$ while holding the persistence
length $\xi$ and chain length $L$
constant. To take the corresponding continuum limit for KWLC, we
will also hold $\zeta$ constant as $\ell\to0$.

We begin by computing the partition functions for the WLC and KWLC
and demonstrating that there
is a continuum limit of the KWLC. These unconstrained partition functions are required for later
computations. For this case, the partition function factors into independent
contributions from each interior vertex.
In the continuum limit ($\kappa\rightarrow \infty$), the partition function
for each vertex in the WLC model is
\begin{equation}\pnlabel{e:Q}
\pnQ \equiv \lim_{\kappa\rightarrow \infty} \int d^2\vec{t}_i\ \exp\left[-\kappa(1-\cos \theta_i)\right] = \frac{2\pi}{\kappa},
\end{equation}
where $\theta_i$ is the polar angle of $\vec{t}_i$  defined using
$\vec{t}_{i-1}$ as the polar axis, that is,
$\cos \theta_i \equiv \vec{t}_{i}\cdot \vec{t}_{i-1}$. The measure
$d^2\vec{t}_i=d(\cos \theta_i)d\phi_i$ denotes solid angle on the unit
sphere. The total discretized partition function for the chain of $N+1$
segments is then
\begin{equation}{\cal Z}\pndis(L) = 4\pi\pnQ^N.\end{equation}
The factor of $4\pi$ reflects one overall orientation integral, for
example the integral over $\vec t_0$.

Similarly, the partition function for a single vertex of the KWLC theory is
\begin{equation}
\pnlabel{dklwcpf}
\pnQ^* \equiv \lim_{\kappa\rightarrow \infty} \int d^2\vec{t}_i
\ \left(\exp\left[-\kappa(1-\cos \theta_i)\right]+\exp\left[-\epsilon\right]\right) =
\pnQ\left(1+2\kappa e^{-\epsilon}\right),
\end{equation}
which we have written in terms of the corresponding WLC quantity
$\pnQ$. The total partition function for the chain of $N+1$
segments is ${\cal Z}^*\pndis(L) = 4\pi(\pnQ^{*})^N$.

In the small segment length limit, \eref{dklwcpf} shows that the probability of a vertex being kink-like is $2\kappa e^{-\epsilon}$.
Therefore the probability of kinking per unit length (for this
unconstrained situation) is
\begin{equation}
\pnlabel{zeta}
\zeta \equiv \frac{2\xi}{\ell^2}e^{-\epsilon}= \frac{4\pi}{\ell \pnQ}e^{-\epsilon},
\end{equation}
where we have eliminated the bending spring constant, $\kappa$, in
favor of the persistence length, $\xi=\kappa\ell$. In order to recover a
sensible continuum limit, we will hold the parameter $\zeta$ constant as we
take the segment length to zero. Note that we recover the WLC theory
when we set $\zeta\to0$. \pn{In later sections we will discuss a
formal ``zero temperature'' limit, in which}{Typically in physical
systems simple enough that the energetic and entropic
contributions are separable, it would not be convenient to take
kink density as the fundamental parameter in the theory since this
parameter would be temperature dependent. But since the kinking
behavior and energy reflect collective phenomena, it is not useful to make this
distinction. Furthermore the persistence length (or elastic
modulus) is itself a composite of energetic and entropic
contributions already and therefore it too is temperature
dependent. It is important to  make this distinction now, because
later on we formally discuss a ``zero temperature'' limit which
is the limit in which} simple mechanics (no thermal
fluctuations) describes the physics. This limit is a useful
intuitive tool, not an experimental prediction of the behavior of
polymers frozen in solution. The ``zero temperature'' limit is
taken treating $\zeta$ as temperature independent, which is
equivalent to either the short rod limit or the large persistence
length limit which we shall use interchangeably.

In the continuum limit, we must remove a divergent constant in  the partition
functions as $N\to\infty$. Thus we define the path integral measure
\begin{equation}
\pnlabel{normalization}
\pnmeas{\vti} \equiv \prod_{i=1}^{N} \frac{d^2\vec t_i}{\pnQ},
\end{equation}
where $\pnQ$ is defined by \eref{e:Q}. Note that unlike the discrete
case, in this measure the starting tangent
vector $\vec t_0$ is \textit{not} integrated, but is instead fixed to
some given $\vti$. The continuum partition function corresponding
to ${\cal Z}\pndis(L)$ is then
\begin{equation}
{\cal Z}(L)\equiv\int\pnmeas{\vti} \ex{-E^{*}} =1.
\end{equation}
With our choice of integration measure, ${\cal Z}(L)$ just equals
one, independent of $L$.

The continuum KWLC partition function is now
\begin{equation}
\pnlabel{KWLCpf}
{{\cal Z}^*}(L) = \lim_{N\rightarrow \infty} \left(1+\frac{\zeta L}{N}\right)^N = e^{\zeta L}.
\end{equation}
The convergence of the partition function assures us that the continuum limit is well defined.
As a consistency check, we now compute the average kink number for the unconstrained chain
\begin{equation}
\left<\kn\right> = \frac{\partial \log {\cal Z}^*}{\partial \log \zeta} = \zeta L,
\end{equation}
which confirms that $\zeta$ is indeed  density of kinks. The expansion of the partition function
in a power series shows that the kink number distribution is also correct.
We will repeat the average kink number calculation several times in the course of this paper for
different constraints to show that
constraining the chain will affect the kink number.

\section{Tangent partition function and propagator}
\pnlabel{tanpart}
\begin{figure}
\includegraphics{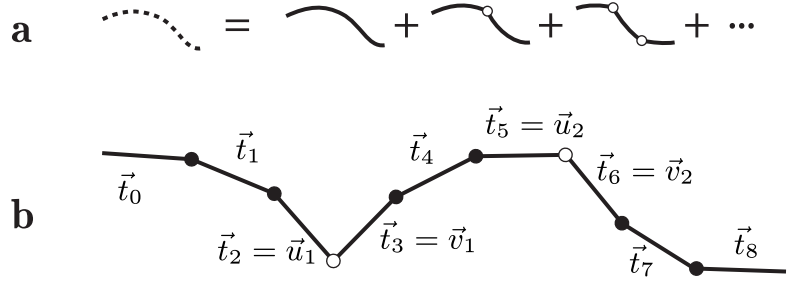}
\caption{\pnlabel{kink.fig.k} {\it a. }  Diagrammatic representation of the kink expansion
for the tangent partition function. The dashed curve represents the KWLC
theory and the solid curves represent the WLC theory. It is convenient to collect the
terms by kink number as shown. {\it b.~}~Detail of the two-kink term,
showing the relation to the underlying discrete model.
$\vum{i}$ and $\vvm{i}$ are the tangent vectors flanking kink number
$i$.}
\end{figure}
In this section we  compute the tangent partition function
and propagator by using a method symbolized in \fref{kink.fig.k}a. By tangent
partition function ${\cal Z}(\vtf,\vti,L)$ we mean the
partition function with the initial and final tangents
constrained (\eref{wlctpf} below).
Dividing the tangent partition function by the
unconstrained partition function ${\cal Z}(L)$ gives the probability
density $H(\vtf,\vti,L)$ for the final tangent vector, given the
initial tangent. We will refer to $H$ as the
normalized tangent partition function, or propagator.

Most of the kink-related physics of the KWLC theory
can be understood qualitatively from the tangent partition
function. Furthermore, the computation of the tangent partition
function is more transparent than the analogous spatial
computation in which the end-to-end distance is constrained along
with the initial and final tangents. The tangent partition function for WLC is defined as
\begin{equation}
\pnlabel{wlctpf}
{\cal Z}(\vtf,\vti;L) \equiv \int\pnmeas{\vti} \ e^{-E} \
\delta^{(2)}[\,\vt_N-\vtf\ ],
\end{equation}
where the path integral is regularized as described above (eqn \ref{normalization}).
Due to the tangent constraint, the partition
function no longer factors into independent vertex contributions.
The lower limit on the integration
denotes that the initial tangent $\vt_0$ is held equal to $\vti$; the
final tangent $\vt_N$, is set to
$\vtf$ by the delta function. We integrate (or sum) over the infinite set of intervening tangents in
order to generate the partition function.
In this regularization scheme, the tangent partition function and tangent propagator are identical
\begin{equation}
\pnlabel{wlcp}
H(\vtf,\vti;L) = {\cal Z}(\vtf,\vti;L),
\end{equation}
since with our conventions the unconstrained WLC partition function is
one.  However, we will see that this convenient identity does not hold for
the KWLC: $H^*\not={\cal Z}^*$.

While the direct evaluation of the path integral in eqn \ref{wlctpf} is difficult,
it is well known that this calculation is equivalent to finding the quantum-mechanical
propagator for a particle on the unit sphere\cite{FeynmanHibbs,Grosberg}.
The tangents correspond to position, arc length corresponds to imaginary time, and
persistence length corresponds to mass. Thus, the tangent partition function is
\begin{eqnarray}
{\cal Z}(\vtf,\vti;L) = \left<\vtf\,\right|e^{-{\cal H}L}\left|\vti\right>,
\end{eqnarray}
where the Hamiltonian operator is defined as
\begin{equation}
{\cal H} \equiv \frac{\vec{p}^{\ 2}}{2\xi},
\end{equation}
where $\vec{p}^{\ 2}$ is the Laplace operator on the unit sphere. The Hamiltonian is
diagonal in the angular momentum representation so the tangent
partition function for WLC can be expressed as
\begin{equation}\pnlabel{e:tpfWLC}
{\cal Z}(\vtf,\vti;L) = \sum^{\infty}_{l=0}\sum_{m=-l}^{l}
Y^m_l(\vtf{\, }) Y_l^m(\vti)^*C_l(L).
\end{equation}
In this expression, the $Y_l^m$'s are the Spherical Harmonics and the
coefficients $C_l$ are
\begin{equation}
\pnlabel{cl}
C_l(L) = \exp\left[-\frac{l(l+1)L}{2\xi}\right].
\end{equation}
It can easily be shown that this partition function has the required
normalization by summing over the final tangent to recover ${\cal Z}(L) =1$.

To compute the tangent partition function for KWLC, we proceed
with the path integral in exactly the same fashion, setting the
initial tangent, integrating over an infinite set of intervening
tangents, and summing over the state vectors:
\begin{equation}
\pnlabel{etp}
{\cal Z}^*(\vtf,\vti;L) = \sum_{\{\sigma_1,\ldots,\sigma_N\}} \int\pnmeas{\vti}\ e^{-E^*}\
\delta^{(2)}[\,\vt_N-\vtf\ ].
\end{equation}
It is now convenient to collect the terms in contributions with a
fixed number $\kn$ of kinks and then express the result in the continuum
limit.

The first step in going from the definition of the discrete KWLC
tangent partition function to the continuum limit is to
reorganize the sum over $\{\sigma_n\}$ as a sum over the number of
kinks $\kn $. Each term of this sum is in turn a sum over the
positions $n_i$ of the kinks, for $i=1,\ldots \kn $. The only subtlety here is
introducing the correct limits on the sum to avoid over counting
the kink states. The last kink can be chosen at any arc-length
location, but additional kinks must always be chosen with smaller
arc-length values than the following kink. This method is more
convenient than introducing ``time ordering'' and a factor of
$1/\kn !$ to explicitly remove the over counting as is commonly done
in the Dyson expansion for time-dependent quantum perturbation
theory\cite{sakurai}.

The next step is to replace the kink position sums with integrals
over the position of the kinks as
\begin{equation}
\sum_{n_i=i}^{n_{i+1}-1} \rightarrow \int_0^{L_{i+1}}\frac{dL_i}{\ell},
\qquad i=1,\ldots \kn \ ,
\end{equation}
where $L_i \equiv n_i\ell$ and we take $L_{\kn +1}=L$. The structure of the arc length integrals is that of a series of
convolutions\cite{arfken}, which we write symbolically as $\otimes$.
For example, if $F(L)$ and $G(L)$ are two functions, then
\begin{equation}
(F\otimes G)(L)\equiv\int_{0}^L\,dL_1\ F(L-L_1)G(L_1)\ .
    \end{equation}
In the intervals between kinks, the chain is described by the WLC
energy function. We can therefore replace each partial path integral with a WLC propagator.

For every kink, there is one factor of $\pnQ\inv$ that
has been introduced by the path integral normalization (eqn \ref{normalization})
but is not absorbed by the
definition of the WLC propagator (eqns \ref{etp} and \ref{wlcp}).
The $\kn $ factors of $\ell\inv$, $e^{-\epsilon}$, and $\pnQ\inv$ can now be
written as $(\zeta/4\pi)^\kn $ (see eqn
\ref{zeta}). Defining ${\cal Z}^{*}=\sum_{\kn }{\cal Z}^*_\kn $,
the terms in the kink-number expansion can thus be written (compare \fref{kink.fig.k})
\begin{equation}
\pnlabel{kwlcpfcont} {\cal Z}^*_\kn (\vtf,\vti;L) = \zeta^\kn \int
\prod_{j=1}^{\kn } \frac{d^2\vu_{j}d^2\vv_{j}}{4\pi}\
\underbrace{\left(H(\vtf,\vv_\kn )\otimes
H(\vu_m,\vv_{m-1})\otimes \cdots\otimes H(\vu_1,\vti)\right)}_{\kn
+1}(L),
\end{equation}
The $2\kn $ angular integrations are over the incoming
($\vu_i$) and outgoing ($\vv_i$) tangents of the
$\kn $ kinks.

\Eref{kwlcpfcont} has a very simple interpretation. The
probability of creating a kink between $L$ and $L+dL$ is just
$\zeta dL$. We then sum over all possible configurations being
careful to choose the integration limits so as not to over count
the kink states. At each kink, all orientational information is
lost, so that only tangent independent terms of the propagator
contribute (those with angular quantum number $l=0$).

To compute the contour length convolution of propagators, it is
convenient to work with the contour length Laplace transformed
propagators $\tilde H$ (eqn \ref{laplace}). We shall denote the
contour length Laplace transformed functions with a tilde and use
the variable $p$ as the arc length Laplace conjugate variable.
Although we could avoid Laplace transforming the partition function
at this juncture, we use this method presently because it is analogous
to our later computation of the spatial propagator. By the well
known Faltung theorem (eqn \ref{ftlt}), the convolution of
propagators is  just the product of Laplace transforms.
Therefore, in terms of the transformed WLC propagators, the $\kn $
kink KWLC Laplace transformed partition function is
\begin{equation}
\pnlabel{nke}
\tilde {\cal Z}^*_\kn (\vtf,\vti ;p) = \zeta^\kn  \begin{cases} \tilde H(\vtf,\vti ;p), & \kn =0  \\
                                            \tilde C_0^{\kn +1}(p)/4\pi,  &
                                            \kn >0\ .\\
\end{cases}
\end{equation}
To derive \eref{nke}, note that \eref{e:tpfWLC} gives the WLC tangent propagator
summed over the initial tangent as $C_0(L)$, which equals 1 from
\eref{cl}. The corresponding
Laplace transform is just $\tilde C_0(p) = 1/p$.

The $\kn $ kink contributions to the KWLC transformed tangent
partition function can now be summed exactly (i.e. ${\cal Z}^{*}=\sum_{\kn }{\cal Z}^*_\kn $) because they form a
geometric series, resulting in
\begin{equation}
\tilde {\cal Z}^*(\vtf,\vti ;p) = \tilde H(\vtf,\vti ;p) +
\frac{1}{4\pi}\frac{\zeta \tilde C_0^2(p)}{1-\zeta \tilde C_0(p)}\ .
\end{equation}
The $\kn >0$ kink terms clearly contribute no tangent dependence.
The inverse Laplace transform can now
be computed without complications, giving the exact KWLC tangent partition function
\begin{equation}\label{e:xkpfl}
{\cal Z}^*(\vtf,\vti ;L) = H(\vtf,\vti ;L) + \frac{e^{\zeta L}-1}{4\pi}.
\end{equation}

Alternatively, we could have derived \eref{e:xkpfl} by noting that
the KWLC model is mathematically equivalent to a Quantum
Mechanical system whose Hamiltonian is diagonal in the angular
momentum representation:
\begin{equation}
{\cal H}^* = -\zeta\ \left|0,0\right>\otimes\left<0,0\right| + {\cal
H}\ .
\end{equation}
Here $\left|\, l,m\right>$ is the state with angular momentum quantum numbers $l$ and $m$ and
$\cal H$ is the Hamiltonian operator for the WLC.
The only change to the theory is a ``ground state energy'' shift equal
to $-\zeta$.

The KWLC tangent propagator and its Laplace transform can now be
evaluated using \eref{KWLCpf}:
\begin{eqnarray}
\pnlabel{kwlctp}
H^*(\vtf,\vti ;L) &=& \frac{{\cal Z}^*(\vtf,\vti ;L)}{{\cal Z}^*(L)} =  e^{-\zeta L} \left[H(\vtf,\vti ;L) + \frac{e^{\zeta L}-1}{4\pi}\right], \\
\tilde{H}^*(\vtf,\vti ;p) &=& \tilde H(\vtf,\vti ;p+\zeta) + \frac{\zeta}{4\pi p(p+\zeta)}.
\end{eqnarray}
\Fref{propenergy}a compares the KWLC tangent propagator to the WLC theory with an
illustrative value $\zeta=0.01/\xi$. The two theories appear
indistinguishable, and in fact we will find that many, but not all, predictions of
the models are essentially the same in this parameter regime.

In principle since the propagator is known exactly, everything in
the theory can now be computed. Of course this is an exaggeration
since, even though the tangent propagator for WLC has long been
known, only recently have the exact expressions for the
transformed spatial propagator been derived\cite{Andy,SS}. The
free energy of the chains for both theories have the canonical
relation with their respective partition functions:
\begin{equation}\label{e:dfF}
F(\theta;L) = -\log {\cal Z}(\vtf,\vti ;L),
\end{equation}
where we have explicitly written the free energy in terms of the deflection angle defined by the
dot product of the initial and final tangents: $\cos \theta = \vti\cdot\vtf$. Up to this point
we have written the partition function and propagator as explicit functions of both the initial and the final
tangent but the rigid body rotational invariance of the energy implies that these functions depend only on
the deflection angle. To express any quantity in terms of the deflection angle, we set the initial tangent to
be the unit vector in the $z$ direction and the final tangent to be the unit vector in the radial direction.
$\theta$ now assumes its canonical definition in spherical polar coordinates.

\Fref{propenergy}b compares the free energies of WLC and KWLC. Despite
the similarity of propagators (\fref{propenergy}a), the free energies
are quite different.
To understand the significance of this free
energy, we imagine discretizing the chain at some segment length
$\ell$. The free energy $F(\theta;\ell)$ gives us the effective
constitutive relation for single-state torsional springs in this
new discretized theory. As depicted is fig \ref{propenergy}b, the
potential energy of these springs is initially quadratic in deflection,
but saturates due to kink formation.
\begin{figure}
\includegraphics{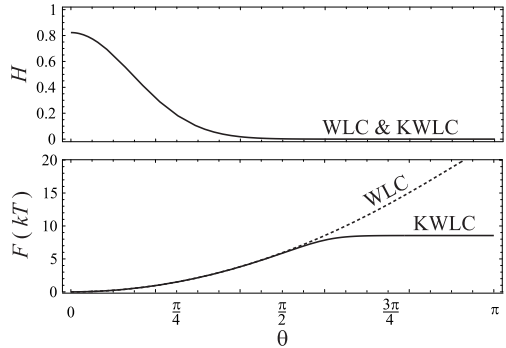}
\caption{\pnlabel{propenergy} The tangent propagator and the tangent
free energy as functions of the deflection angle for the illustrative
values $L = 0.2\xi$
and $\zeta\xi = 10^{-2}$. The solid curves are KWLC and the dashed
curves are WLC with the same value of $\xi$. In the
absence of kinking, the WLC distribution ($H$) is essentially zero
away from small deflection. For the small value of $\zeta$ chosen
above, WLC and KWLC are indistinguishable in the top panel. The
presence of kinks adds a background level to the propagator which
is independent of $\theta$, but is thermally inaccessible---too
small to distinguish from zero in the top panel, but is visible in
the free energy in the lower panel. The tangent free energy gives
an intuitive picture of the system interpreted as as single-state
system with an effective bending modulus which saturates due to
kinking. Most thermally driven experiments measure the polymer
distribution as it is pictured in the top panel and are therefore
insensitive to the high-curvature constitutive relation. But
experiments which do probe this regime, short-contour-length
cyclization for example, will be extremely sensitive to the
difference between the theories due to the large free energy
difference at large deflection.  }
\end{figure}

\subsection{Moment-Bend \& kink number}
\begin{figure}
\includegraphics{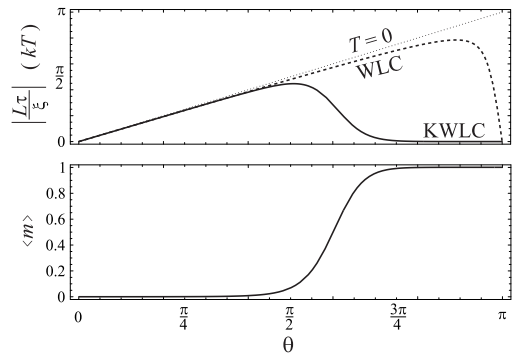}
\caption{\pnlabel{momentbend} The bending moment $\moment$ and average kink
number $\langle \kn\rangle$ as functions of the tangent deflection angle for illustrative
values $L =
0.2\xi$ and $\zeta\xi =  10^{-2}$. The solid curves are KWLC and
the dashed curves are WLC with the same bend persistence length.
At small $\theta$, the normalized bending moment exhibits a linear
spring dependence and the chain is unkinked. The limiting linear
behavior of the short rod limit is the dotted curve, labeled $T=0$
corresponding to the mechanical limit of WLC. For large
deflection, the chain kinks and the moment drops to zero. This
correspondence between kinking and the moment is clearly
illustrated in the short length limit depicted above. }
\end{figure}
To understand the interplay between chain kinking and deflection,
it is helpful to explicitly compute the relation between the
deflection angle and the restoring moment (torque), as well as
computing the average kink number. Here we ask the reader to
imagine a set of experiments analogous to force-extension but
where the moment-bend constitutive relation is measured. We
compute the constitutive relation in the usual way in terms of the
deflection angle $\theta$
\begin{equation}
\moment(\theta;L) \equiv -\frac{\partial}{\partial \theta} F (\theta;L),
\end{equation}
where $F(\theta;L)$ is the tangent free energy and $\theta$ is the
deflection angle. In terms of the WLC bending moment, \eref{e:xkpfl} shows
that the moment for KWLC has a very simple form:
\begin{equation}
\pnlabel{KWLCm}
\moment^*(\theta;L) = \moment(\theta;L)\frac{{\cal Z}(\theta;L)}{{\cal Z}^*(\theta;L)},
\end{equation}
where $\moment$ is the WLC moment and ${\cal Z}$ and ${\cal Z}^*$ are the tangent partition functions
for WLC and KWLC, respectively. The moment is plotted as a
function of deflection in fig \ref{momentbend}.
For short chains, the small deflection moments of the two theories
initially coincide. But as the deflection increases, there is a transition, corresponding to the
onset of kinking, where the moment is dramatically reduced to nearly zero.
In eqn \ref{KWLCm}, this transition is clear from the ratio of the partition functions.
Remember that the KWLC partition function is the sum of the WLC partition function and
the $\kn >0$ kink partition functions. Before the onset of kinking, the WLC and KWLC partition
functions are equal since
the kinked states do not  contribute significantly to the partition function.
For large deflection, the KWLC partition function is
kink dominated and therefore the ratio in eqn \ref{KWLCm} tends to zero.
Physically, once the chain is kink dominated, the moment must be zero since the kink energy is independent
of the kink angle. At zero temperature, the moment would be zero, but fluctuations in which the
chain becomes unkinked cause the moment to be nonzero. We discuss this effect in more detail below.

To explicitly see that the reduction in the moment corresponds
to kinking, we compute the average kink number as a function of deflection:
\begin{eqnarray}
\left<\kn\right>(\theta;L) = -\frac{\partial F^*}{\partial \log \zeta} =
\frac{\zeta L}{4 \pi H^*(\theta;L)}\ ,
\end{eqnarray}
which is depicted in fig \ref{momentbend}. Note that when we
remove the tangent constraint, we again find that the average kink
number is $\zeta L$. When the chain is constrained, the
enhancement factor is proportional to $(H^*)\inv$. Note that this
implies that the kink number will be reduced when the tangents are
constrained to be aligned and enhanced when the chain is
significantly bent. In fig \ref{momentbend}, the kink-induced
reduction in the moment can be seen to correspond to the rise of
the kink number from zero to one kink.

We will now compare these exact results to the mechanical or ``zero temperature'' limit.
This regime is equivalent to the large persistence length limit, where we can write the
partition function concisely as
\begin{equation}
\pnlabel{smallLlimit}
\lim_{L\rightarrow 0} {\cal Z}^*(\vtf,\vti ;L) = \frac{1}{4\pi}
\left[\frac{2\xi}{L} \exp\left(-\frac{\xi}{2L}\theta^2\right) + \zeta
L\right]\ ;
\end{equation}
the WLC limit is recovered for $\zeta=0$.
In the short length limit, the moment of the WLC chain is simply linear in deflection: $\moment=-\theta \xi/L$.
This moment is also plotted in fig \ref{momentbend}. Even without the complication of kinking, there is
already one interesting feature of the exact WLC moment-bend constitutive relation
which needs explanation. For large deflection, the linear
relation already fails in the WLC model! This is a thermal effect which is best understood by going to the
extreme example of deflection $\theta=\pi$. For any configuration, the contribution of a chain reflected
through the axis defined
by the initial tangent will make the partition function symmetric about
$\vtf=-\vti$. This implies that
the bending moments from these chains cancel. Away from $\theta=\pi$, the cancellation is no longer exact. In the
mechanical limit, this effect is present but localized at $\theta = \pi$ due to the path degeneracy.

In the mechanical limit, kinking is always induced by bending and
at most one kink is nucleated. In this limit, the KWLC bend-moment
can be rewritten in terms of the kink number
\begin{equation}
\moment^*(\theta;L) = \moment(\theta;L)(1-\left<\kn\right>),
\end{equation}
where the kink number is just the Heaviside step function,
\begin{equation}
\left<\kn\right>(\theta,L) = \Theta_H[\theta-\theta_0],
\end{equation}
around a critical deflection angle
\begin{equation}
\theta_0 \equiv \left[\frac{2L}{\xi} \log \frac{2\xi}{\zeta L^2} \right]^{1/2}.
\end{equation}
For deflection less than the critical deflection, the kink number is zero and the moment is
given by the WLC moment. At the critical deflection angle, there is an abrupt transition to
the kinked state with kink number one and the moment zero. Precisely at the critical angle the free
energy of the kinked state and the elastically bent state are equal. Note that we have
not discussed dynamics and have assumed that the system is in equilibrium, not
kinetically trapped.

The behavior of the KWLC theory for short contour lengths is
nearly what one would expect from mechanical intuition. Bending of
the chain on short length scales induces a moment which is
initially linearly dependent on deflection. When the chain is
constrained to a large deflection angle, kinking is induced and
the response of the chain to deflection is dramatically weakened.
In the mechanical limit, once kinking is induced, the moment is
zero but for finite size rods, the ability of the chain to
fluctuate between the kinked states and unkinked states blurs the
dramatic zero-temperature transition between the kinked and
unkinked bend response.

Our discussion here has focused
principally on developing an intuition for the short chain limit.
From an experimental perspective, it is difficult to measure the
moment-bend relation directly as we have described, especially for
short chains. While single molecule AFM experiments might probe
this relation, most of the information about the moment-bend
constitutive relation comes from indirect measurements of
thermally-induced bending. For example light scattering,
force-extension, and cyclization experiments are all measures of
thermally induced bending. As we shall explain, only cyclization
experiments with short contour length polymers are sensitive to
the high curvature regime of the moment-bend constitutive
relation. For the most part, these thermally driven bending
experiments are only sensitive to the thermally accessible regime
of the moment-bend constitutive relation which corresponds to
small curvature and therefore small deflection on short length
scales, a regime that is very well approximated by linear
moment-bend constitutive relation. For long chains, the initial
linear response is weaker implying that the kinking transition is
less pronounced. In fact we shall see in the next section that for
some of these indirect measurements of the low curvature regime of
the constitutive relation, the effect of the kinking will be
indistinguishable from the linear elastic response.

\subsection{Persistence length}
\pnlabel{ps} Since many polymer characterization experiments are
most sensitive to the thermally-accessible weak-bending regime, it
is clearly of interest to determine whether kinking changes this
low-curvature physics. Intuitively, we have already argued that,
at least for small kink densities, many properties of the polymer
that do not explicitly probe the highly bent structure will remain
essentially unchanged. In this section, we will derive a number of
exact results that show that the effects of kinking can be
described by a renormalization of the persistence length of WLC
theory for some bulk features of the polymer distribution,
regardless of the magnitude of $\zeta$.

The tangent-tangent correlation must be a decreasing exponential due to
multiplicativity\cite{PhilBook,Grosberg}
and therefore we can discuss the decay length. The tangent persistence is
\begin{eqnarray}
\left<\vec{t}_\Delta  \cdot \vec{t}_0\right> = e^{-(\zeta+\xi^{-1})\Delta },
\end{eqnarray}
which can be computed by examining the limit of small $\Delta$ and
applying the tangent propagator (eqn
\ref{kwlctp}).
Since this result is identical to the WLC result except for the decay constant, we
introduce the effective persistence length
\begin{equation}
\pnlabel{renorm}
\xst \equiv \left(\xi^{-1}
+\xk ^{-1}\right)^{-1},
\end{equation}
where the kink length is defined as
$\xk  \equiv \zeta^{-1}$. The form of this effective persistence
length is not surprising since a roughly analogous effect is
observed adding two linear springs together or from the
combination of static and dynamic persistence
length\cite{trif87a,nels98bx,wiggins3}. This tangent persistence
result immediately implies that an analogous exact renormalization
occurs for both the mean squared end distance
\begin{eqnarray}
\pnlabel{ased}
\left<R^2\right>_{\rm KWLC} = \left[\left<R^2\right>_{\rm
WLC}\right]_{\xi\rightarrow  \xst} =2L \xst-2{ \xst}\left(1-e^{-L/ \xst}\right)
\end{eqnarray}
and the radius of gyration
\begin{eqnarray}
\left<R^2_g\right>_{\rm KWLC} = \left[\left<R^2_g\right>_{\rm
WLC}\right]_{\xi\rightarrow  \xst} =  \frac{L{ \xst}}{3}-{
\xst}^2+\frac{{2 \xst}^3}{L}-\frac{2{ \xst}^4}{L^2}\left(1-e^{-L/{
\xst}}\right),
\end{eqnarray}
since these result are simply integrations of the tangent persistence\cite{YamakawaBook}.
In experiments sensitive only to the radius of gyration (static scattering for small wave number) or the average
square end distance (force-extension in the small force limit), the measured
persistence length of the KWLC theory will be the effective persistence
length, $ \xst$, regardless of the
magnitude of $\zeta$.
In most systems of physical interest, the kink length is much larger than the bend persistence
length implying that, even if the bend persistence were independently measurable, the difference
between the effective persistence length and the bend persistence length would be very small.
In other words, the loss of tangent persistence due to kinking is negligible compared with the
loss due to thermal bending since kinks are rare on the length scale of a persistence length.

The tangent persistence corresponds to the first moment of the tangent propagator. Clearly the
renormalization we have discussed fails for higher order moments! At least in principle
it is therefore possible to determine the bend persistence from higher order moments of the distribution.
From an experimental perspective this corresponds to scattering experiments at large wave number,
force extension for large force, or cyclization experiments for short contour length.
To predict the effects of kinking in these experiments, we must compute the spatial
propagator (the spatial distribution function).

\section{Tangent-spatial and spatial propagators}
\pnlabel{tssp} The spatial propagator $K(\vec{x};L)$ is defined as
the probability density of end displacement $\vec{x}$ for
a polymer of contour length $L$. Similarly, the tangent-spatial
propagator $G(\vec{x};\vtf,\vti ;L)$ is defined as the
probability density of end displacement $\vec{x}$ with final
tangent $\vtf$, given an initial tangent $\vti$, for a
chain of contour length $L$. Although in principle the theory is
solved once the tangent propagator is known, the moments of the
spatial propagator, or spatial distribution function, are more
experimentally accessible than the tangent propagator.  In particular, the
$J$ factor measured in cyclization experiments, the force-extension characteristics,
and the structure factor measured in scattering experiments are
all more directly computable from the propagators $G$ and $K$. In
this section, we first compute the spatial propagator and then
discuss its application to experimental observables.

Following our computation of the tangent propagator, we
compute the tangent-spatial and spatial partition functions. Our
solution relies on the same Dyson-like expansion of the partition
function in the kink number as was exploited to compute the
tangent partition function. The only added complication is that,
in addition to the arc length convolution, we must also compute
convolutions over the 3d spatial positions of the kinks. By going to
the Fourier-Laplace transformed propagator, the convolutions
again become products and the $\kn $ kink contributions can be summed
exactly. Unfortunately the exact results of this computation will
only be found analytically up to a Fourier-Laplace transform, in
part because the WLC theory itself is only known analytically in
this form\cite{Andy,SS}.

We begin by writing the tangent-spatial partition function
for the KWLC theory in an form analogous to the tangent partition function in eqn \ref{etp}:
\begin{eqnarray}
\pnlabel{kwlcpfd}
{\cal Z}^*(\vec{x}; \vtf,\vti ; L) &=&
\sum_{\{\sigma_1,\ldots,\sigma_N\}} \int\pnmeas{\vti}\
e^{-E^*}\delta^{(2)}[\,\vt_N-\vtf\ ]\delta^{(3)}[\,\vx_{N+1}-\vx\ ],
\end{eqnarray}
where $\vec{t}_0$ is the initial tangent vector. The additional
spatial Dirac delta function in the equation sets a spatial
constraint for the end displacement; in this expression,
$\vx_{N+1}\equiv\ell\sum_{n=0}^{N}\vt_n$. We will again collect the
terms in this sum by kink number $\kn $. In the intervals between
kinks, we again introduce the WLC propagator, but this time we use
the tangent-spatial propagator $G$, defined by an expression
analogous to eqn \ref{kwlcpfd}, but with $E$ in place of $E^*$.

Because we have normalized the
unconstrained WLC partition function such that ${\cal Z}\equiv 1$,
the tangent-spatial partition function and propagator are
identical. It is convenient to introduce the WLC spatial
propagator
\begin{eqnarray}
\pnlabel{spe}
K(\vec{x};L) \equiv  \frac{1}{4\pi} \int\, d^2\vt_1 d^2\vt_2\ G(\vec{x};\vec{t}_1,\vec{t}_2;L),
\end{eqnarray}
where we sum over the final tangent and average over the initial
tangent to derive the spatial probability density. We also introduce the one tangent summed
tangent-spatial propagator
\begin{eqnarray}\label{e:gprime}
G'(\vec{x},\vec{t};L) =  \int d^2\vt_1\ G(\vec{x};\vec{t},\vec{t}_1;L),
\end{eqnarray}
which will allow us to concisely express intermediate results.
Finally for economy of notation, we write the convolutions over both the spatial position and
arc length symbolically with $\otimes$, generalizing the notation
introduced in \sref{tanpart}.

The $\kn >0$ kink KWLC tangent-spatial partition function can be written in terms of the WLC
propagators:
\begin{equation}
{\cal Z}^*_\kn (\vec{x};\vtf,\vti ;L) = \frac{\zeta^\kn }{4\pi}\left(
G'(\vtf) \otimes \left[K\otimes\,\right]^{\kn -1} G'(\vti)
\right)(\vx,L)\ ,
\quad \kn >0\ .
\end{equation}
We now introduce the WLC Fourier-Laplace transforms of
the propagators $G'$ and $K$. We denote the transformed functions
with a tilde. The Laplace conjugate of contour length $L$ is $p$
and the Fourier conjugate of the end displacement $\vec{x}$ is the
wave number $\vec{k}$. The Faltung theorem (Eqs \ref{ftft} and
\ref{ftlt}) allows us to replace the spatial-arc length
convolutions with the products of the Fourier-Laplace transformed
propagators. The $\kn $ kink KWLC transformed partition function is
\begin{equation}
\tilde {\cal Z}^*_\kn (\vec{k};\vtf,\vti ;p) = \zeta^\kn  \begin{cases} \tilde
G(\vk,\vtf,\vti;p), & \kn =0\\
\tilde G'(\vk,\vtf;p) \tilde K^{\kn -1}(\vk;p) \tilde G'(\vk,\vti;p)/4\pi, & \kn >0,\end{cases}
\end{equation}
which is analogous to eqn \ref{nke} for the tangent propagator.

As before, the transformed $\kn $ kink contributions can be summed exactly in a
geometric series. Abbreviating the notation somewhat, the resulting
tangent-spatial transformed partition function becomes
\begin{equation}
\pnlabel{tsp}
\tilde {\cal Z}^*(\vec{k};\vtf,\vti ;p) = \tilde G + \frac{\zeta \tilde
G' \tilde G'}{4\pi \left(1-\zeta \tilde K\right)}\ .
\end{equation}
We can also derive the KWLC transformed spatial partition function by averaging
over the initial tangent and summing over the final tangent. Applying
the definition in \eref{spe} gives
\begin{equation}
\tilde {\cal Z}^*(\vec{k};p) = \frac{\tilde K(\vk;p)}{1-\zeta \tilde
K(\vk;p)}\ .
\end{equation}
To compute the KWLC spatial and tangent-spatial propagators, we divide the
constrained partition functions by the unconstrained partition function (eqn \ref{KWLCpf}).
The transformed tangent-spatial and spatial propagators are
\begin{eqnarray}
\pnlabel{kwlctsp}
\tilde G^*(\vec{k},\vtf,\vti ;p) &=& {\cal L}{\cal F}\left[\frac{\tilde {\cal Z}^*(\vec{x},\vtf,\vti ;L)}{\tilde {\cal Z}^*(L)}\right] = \tilde {\cal Z}^*(\vec{k};\vtf,\vti ;p+\zeta), \\
\pnlabel{fltsp} \tilde K^*(\vec{k};p) &=& {\cal L}{\cal
F}\left[\frac{\tilde {\cal Z}^*(\vec{x};L)}{\tilde {\cal
Z}^*(L)}\right] = \tilde {\cal Z}^*(\vec{k};p+\zeta),
\end{eqnarray}
where $\cal L$ is the arc-length Laplace transform and $\cal F$ is
the spatial Fourier transform. The transformed WLC spatial
propagator is exactly known\cite{Andy,SS}
\begin{eqnarray}
\pnlabel{kprop}
\tilde K(\vec{k};p) &=& \frac{{\displaystyle1}}{{
\displaystyle P_0+\frac{A_1 \vec{k}^2}{
P_1+\frac{A_2\vec{k}^2}{P_2 +\frac{A_3\vec{k}^2}{\cdots}}
}
}},
\end{eqnarray}
where $A_\pnss$ and $P_\pnss $ are defined
\begin{equation}
A_\pnss  \equiv \frac{\pnss ^2}{4\pnss ^2-1},\ \ P_\pnss  \equiv p +
\frac{\pnss (\pnss +1)}{2\xi}.
\end{equation}
Because the KWLC transformed spatial partition function and
propagator are functions of $\tilde K$, they are also known
exactly. In principle, both $K$ and $K^*$ can be computed by
inverting the transforms numerically. In order to compute the KWLC
tangent-spatial partition function and propagator, the WLC
tangent-spatial propagator, $G$, must also be known. Since $\tilde
G$ is not known analytically, our solution for the tangent-spatial
partition function and propagator are formal. From the perspective
of computing experimental observables, $K^*$ will suffice for
computation of the force-extension characteristic, the structure factor, and
surprisingly, the $J$ factor, despite the tangent constraint in its
definition.

\subsection{Wave number limits}
While we have written the exact transformed propagators for KWLC,
like WLC, these transforms cannot be inverted analytically. It is
therefore useful to examine the exact transformed propagators in
several limits which can be computed analytically. First we
consider the long length scale ($k\rightarrow 0$) limit. We find
that KWLC and WLC are identical apart from the renormalization of
the persistence length (see eqn \ref{renorm}):
\begin{equation}
\pnlabel{smallwavenumberlimit}
\lim_{k\rightarrow 0} {\tilde K^*} = \lim_{k\rightarrow 0} {\tilde
K}_{\xi\rightarrow  \xst} =
\left[p+\frac{1}{3}\frac{k^2}{p+ (\xst)\inv}+\cdots\right]^{-1}.
\end{equation}
By expanding the exponential in the definition of the Fourier transform, it can be shown that this
result is equivalent to showing that $R^2$ is exactly renormalized. In our discussion of the
$J$ factor it will be convenient to consider an even more restrictive limit. We now add the additional
restriction that the chain is long ($p\rightarrow 0$). In this limit we must recover the Gaussian
chain (Central Limit Theorem)
\begin{equation}
\lim_{k,p\rightarrow 0} {\tilde K^*} = \lim_{k\rightarrow 0} {\tilde
K}_{\xi\rightarrow  \xst} =
\left[p+\frac{{ \xst}k^2}{3}+...\right]^{-1},
\end{equation}
which is the transformed Gaussian distribution function for Kuhn
length $2 \xst$. When applicable, the Gaussian
distribution is a power tool due to its simplicity.

The opposite limit is the short length scale
($k\rightarrow\infty$) and short contour length limit
($p\rightarrow\infty$). In this limit WLC and KWLC are identical,
both approaching the rigid rod propagator:
\begin{equation}
\lim_{p,k \rightarrow \infty} \tilde K^* =\lim_{p,k \rightarrow \infty} \tilde K  = \tilde K_{\xi\rightarrow \infty} =
\frac{1}{k} \tan^{-1} \frac{k}{p}.
\end{equation}
The rigid rod spatial propagator describes a polymer that is infinitely stiff. In the limit that we analyze
very short segments of the polymer, both the WLC and KWLC models appear rigid since we have confined our
analysis to length scales on which bending is thermally inaccessible. In this limit, the propagators take a
very simple form which is more tractable than either WLC or KWLC. The rigid rod propagator is useful
when discussing the limiting behavior of the $J$ factor at short contour length and is discussed in
more detail in the appendix.

\subsection{Partition function in an external field and force-extension characteristic}
\begin{figure}
\includegraphics{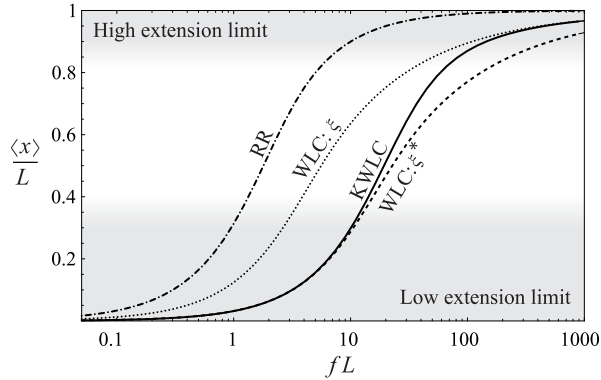}
\caption{\pnlabel{force.ext} Force-extension characteristic for KWLC compared to
WLC and Rigid Rod for $L=4\xi$ and $\zeta\xi = 4$. At low
extension, the force-extension of KWLC (solid curve) approaches
WLC (dashed curve) with a persistence length equal to the
effective persistence length of KWLC. At high extension, the kink
modes are frozen out and the KWLC force-extension characteristic approaches WLC
(dotted curve) with a persistence length equal to the bend
persistence length of KWLC. Rigid Rod (dot dashed curve) has been
plotted for comparison. The extension of Rigid Rod corresponds to
alignment only.  }
\end{figure}
In force-extension experiments, a single polymer molecule is
elongated by a bead in an external field. The average extension of
the polymer is measured as a function of external field strength.
The forces opposing extension are entropic. These entropic forces
are caused by the reduction in the number of available microstates
as the polymer extension is increased. The persistence length
defines the length scale on which the polymer tangents are
correlated. For small persistence length, the number of
statistically uncorrelated tangents is greater, which increases
the size of the entropic contribution to the free energy relative
to the external potential. This deceptively simple physics implies
that a chain with a softer bending modulus acts as a stiffer
entropic spring resisting extension.

To compute the force-extension relation, we must compute the partition function in an
external field $f$ which can be concisely written in terms of the spatial
partition function
\begin{equation}
\pnlabel{fepf}
{\cal Z}_{\vec{f}}\ (L) = \int d^3x \ e^{\vec{f}\cdot\vec{x}} {\cal Z}(\vec{x};L) = \tilde {\cal Z}(i\vec{f};L),
\end{equation}
which is a particularly convenient expression since it is the
Fourier-transformed partition function with the wave number
$\vec{k}$ analytically continued to $i\vec{f}$. Note that this is
the inverse Laplace transform of eqn \ref{tsp}. The average
extension is
\begin{equation}
\left<x(f)\right> = \frac{\partial}{\partial f} \log {\cal Z}_{\vec{f}},
\end{equation}
which may be computed by taking the inverse Laplace transform
numerically. (See \sref{nit} for the numerical method.) The
results are plotted in \fref{force.ext}. In this figure, the KWLC
theory interpolates between two WLC limits at high and low
extension. The low-force limit is clearly related to low wave
number limit (eqn \ref{smallwavenumberlimit}) via an analytic
continuation of the wave number. Therefore KWLC with effective
persistence length $ \xst$ and WLC with persistence length $ \xst$
correspond in the low-extension limit as can be seen in
\fref{force.ext}.

At high force, \fref{force.ext} shows that kinking becomes
irrelevant and the extension of KWLC and WLC both with bend
persistence $\xi$ are identical. In this limit, the chain is confined
to small deflection angles for which the effect of kinking is
negligible, as can be seen in fig \ref{propenergy}. In essence the
kink modes freeze out and measurement of the extension versus
force measures the bend persistence rather than the effective
persistence length of the KWLC polymer chain.

These two regimes imply that in principle the value of $\zeta$ could
be determined by the difference between the persistence length
measured at small and large extension. In practice, this is most
likely not practical. We have purposely chosen an unrealistically
large value of $\zeta$ in \fref{force.ext}, to illustrate clearly the low- and
high-extension limits. In more realistic systems, the difference
between the bend and effective persistence lengths would be small
implying that it would be difficult to detect. Furthermore, at low
extension the effects of polymer-polymer interactions can act to
either increase or decrease the effective low extension
persistence length. At high extension, polymer stretch also acts
to increase the extension at high force most likely obscuring the
effects of the entropy reduction due to the loss of the kink
bending modes. \Fref{philfig} illustrates these remarks.
\begin{figure}
\includegraphics{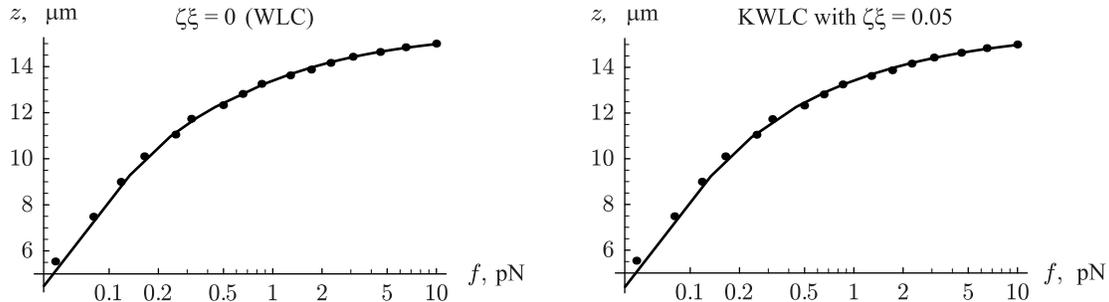}
\caption{\pnlabel{philfig}Left: Semilog plot of the best fit of the WLC model
($\zeta=0$) to experimental data on the force-extension relation of a single
molecule of lambda DNA. Right: Best fit of the KWLC model to the
same data, taking $\zeta\xi=0.05$. The fits are equally good, even
though this value of $\zeta$ is larger than the one that we will
argue fits cyclization data. Thus, force-extension measurements
can only set a weak upper bound on the value of $\zeta$. (Data
kindly supplied by V.~Croquette; see \cite{PhilBook}.)}
\end{figure}
The force-extension characteristic is therefore unlikely to detect the high-curvature
softening induced by kinking.

\subsection{Structure factor}
\begin{figure}
\includegraphics{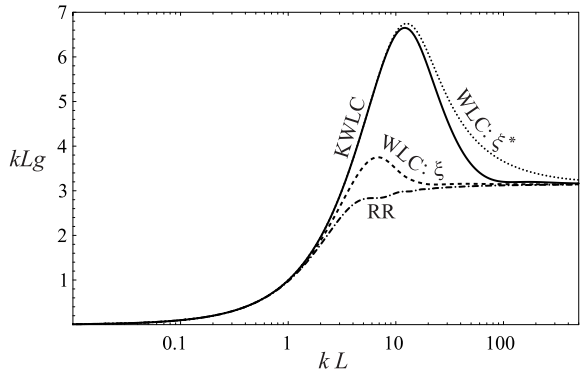}
\caption{\pnlabel{g.struct.renorm} The structure factor and
the role of effective persistence length. The solid curve is the structure
factor for KWLC with contour
length $L=4\xi$, and kink parameter $\zeta = 4/\xi$. For comparison,
we have plotted  the structure factor for WLC of the same contour length for identical bend
persistence lengths (dashed) and identical effective persistence
length (dotted). At short length scales (large wave number) the
KWLC structure factor approaches that for WLC with an identical
bend persistence length. At long length scales (small wave
number), the KWLC structure factor approaches that for WLC with a
persistence length equal to its effective persistence length
$ \xst$. We have also plotted the structure factor for Rigid
Rod (dot dashed curve) for comparison.}
\end{figure}
Another experimental observable used to characterize polymers is
the structure factor, measured by static light scattering,
small-angle X-Ray scattering, and neutron scattering experiments.
Measurements of the structure factor can probe the polymer
configuration on a wide range of length scales. Symbolically the
structure factor is
\begin{equation}
g(\vec{k}) \equiv \frac{1}{L^2} \int_0^Ldsds'\
\left<\exp\left[i\vec{k}\cdot\left(\vec{X}(s)-\vec{X}(s')\right)\right]\right>,
\end{equation}
where $\vec{X}(s)$ is the position of the polymer at arc length
$s$ and we have included an extra factor of the polymer contour
length in the denominator to make the structure factor
dimensionless\cite{Andy}. At high wave number, the structure
factor is sensitive to short length scale physics, whereas the
polymer length and radius of gyration can be measured at low wave
number. The structure factor can be rewritten in terms of the
Laplace-Fourier transformed propagator
\begin{equation}
g(\vec{k}) = \frac{2}{L^2}{\cal L}^{-1}\left[\frac{\tilde K(\vec{k};p)}{p^2}\right],
\end{equation}
where ${\cal L}^{-1}$ is the inverse Laplace transform which can
be computed numerically. (See \sref{nit} for the numerical
method.) As we mentioned above, the leading-order contributions at
small wave vector are the polymer length and the radius of
gyration
\begin{equation}
Lg(k) = L(1+{\textstyle \frac{1}{3}}\vec{k}^2R_g^2+...)
\end{equation}
where we have temporarily restored the length dimension of $g$.
At large $k$, both WLC and KWLC are rod-like or straight which gives an asymptotic limit for
large wave number
\begin{equation}
g(k) \rightarrow \frac{\pi}{Lk},
\end{equation}
since the chain is inflexible at short length scales.

To what extent can scattering  experiments differentiate between
WLC and KWLC? We have already argued that kinking merely leads to
a renormalization of the persistence length for the radius of
gyration, $R_g$, so both theories are identical at the low and
high wave number limits. For the rest of the interval, the
theories do predict subtly different structure factors, but for
small values of $\zeta$, the theories are nearly indistinguishable.
Again, we have chosen to illustrate the structure factor for an
unrealistically large value of $\zeta$, to exaggerate its effect. Like
force-extension measurements, scattering experiments are not
sensitive to the high curvature physics since the signal is
dominated by the thermally accessible bending regime which is
essentially identical to WLC.

\section{Cyclized chains and the $J$ factor}
\pnlabel{Jfactor}
\begin{figure}
\includegraphics{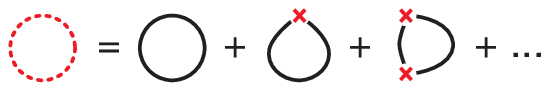}
\caption{\pnlabel{kink.fig} The diagrammatic representation of the kink number expansion for cyclized polymers.
The dashed curve represents the KWLC theory which is the sum of the $\kn $ kink contributions. In the interval
between the kinks, the polymer is described by WLC, represented by the solid curves. For each $\kn $ kink contribution,
we sum over the kink position. In order to meet the tangent alignment conditions for cyclized polymers,
we close the chain at a kink for kink number one or greater. }
\end{figure}
Although the theoretical study of the moment-bend constitutive
relation is straightforward, it is problematic experimentally to
apply a moment and measure the deflection directly on microscopic
length scales. It is typically more convenient to let thermal
fluctuations drive the bending, but as we have discussed above,
experiments which measure thermally-driven bending are typically
not sensitive to the rare kinking events. In contrast, cyclization
experiments, although thermally driven, are sensitive to bending
at any length scale. These experiments measure the relative
concentrations of cyclized monomers to noncyclized dimers. By
choosing the contour length of the monomers, any bending scale may
be studied provided the concentration of cyclized molecules is
detectable. Furthermore, these experiments are typically bulk
rather than single molecule. In fact the data motivating this work
comes from recent DNA cyclization measurements of Cloutier and
Widom\cite{CW} who have shown that the cyclization probability is
$10^4$ to $10^5$ times larger than that predicted by WLC for DNA
sequences with a contour length $L\approx 0.6 \xi$, while
confirming that larger sequences ($L>\xi$) do cyclize at the rate
theoretically predicted by the WLC \footnote{Cloutier and Widom's discussion assumed that the ligase enzyme
used in their experiments acts in the same way when ligating a single DNA or joining two
segments. Although this assumption is standard in the field, it may be
criticized when the length of the DNA loop becomes not much bigger than the
ligase enzyme itself. We believe that effects of this type cannot account for
the immense discrepancy between the measured $J$ factor and that predicted
by the WLC theory.}

In cyclization measurements, the configurational free energy is
isolated in the $J$ factor which is ratio of the cyclization
equilibrium constant to the dimerization equilibrium
constant\cite{JacobsonStockmayer}. This ratio eliminates the
dependence on the end-end interaction free energy. For non-twist
storing polymers, the $J$ factor is proportional to the
tangent-spatial propagator\cite{JacobsonStockmayer}
\begin{equation}\pnlabel{e:dfJ}
J = 4\pi G(0;\vec{t},\vec{t};L),
\end{equation}
which is the concentration of one end of the polymer chain at the other
($\vec{x}=0$) with the correct tangent alignment. The  factor of
$4\pi$ is due to the isotropic angular distribution of monomer in
free solution. Our analysis will neglect additional complications
relevant to the study of real DNA. First, in DNA the twist must
also be aligned, which requires the use of a variant of WLC,
Helical Wormlike Chain\cite{YamakawaBook}. This additional
constraint modulates the $J$ factor with a 10.5 bp period equal to
the helical repeat. Our interest here is in the value of the $J$
factor averaged over a helical repeat for which the effects of
twist can be roughly ignored\cite{YamakawaBook}. A second
complication in real DNA is sequence dependent
prebending\cite{mann96a,zhan03a}. We argue elsewhere that
prebending effects alone cannot explain the high cyclization rates
observed for short DNA\cite{wiggins2}; in this paper we focus
instead on kink formation.
\par Although cyclization
experiments are fairly straightforward, extracting mechanical
information from the results poses a difficult theoretical problem
due to the combination of tangent and spatial constraints. In
fact, there is no exact analytic expression for the $J$ factor in
the WLC theory; the following sections and appendices  will
develop the numerical methods we need.

\subsection{The looping $J$ factor}
\begin{figure}
\includegraphics{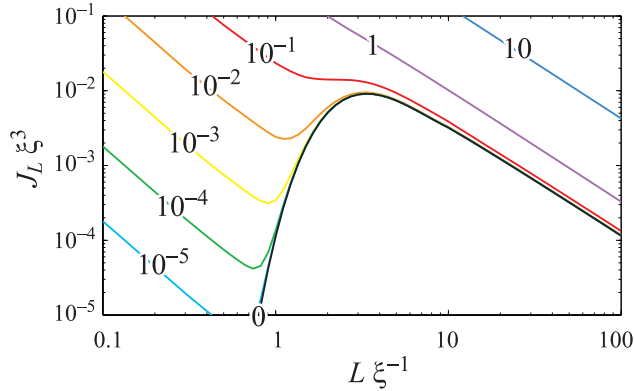}
\caption{\pnlabel{jp}The KWLC looping $J$ factor, $J^*_L$,
as a function of contour length plotted for various values of the kinking
parameter $\zeta$. The
numbers labeling the curves indicate the value of the
dimensionless quantity $\xi \zeta$. WLC is the curve labeled 0. For large contour length $L$, the effect of kinking
can be accounted for by computing $J_L$ for the effective
persistence length, $ \xst$. But as the contour length
shrinks to a persistence length, the effect of kinking becomes
dominant, even for small $\zeta$. At short contour length
the looping $J$ factor is one kink dominated and diverges in
contrast to the WLC looping $J$ factor which  approaches zero
precipitously for short contour length. }
\end{figure}
Due to these computational complications, we shall initially
dispense with the tangent alignment condition and compute a
modified $J$ factor that is relevant for processes that do not fix
the tangents of the chain. For example some protein-DNA complexes
exhibit a  behavior that is believed to be better
represented by looping (free end tangents) than cyclization (end tangents
aligned) \cite{Zhang}. We define the looping $J$ factor as the ratio
of the looping to the dimerization equilibrium constants. The KWLC
looping $J$ factor, $J^*_L$, can be written in terms of the spatial
propagator as
\begin{equation}
J_L^* = K^*(0;L),
\end{equation}
which can be interpreted as the concentration of one end at the other. We have again neglected the
effect of twist.
In this case \pn{the explicit $4\pi$ in \eref{e:dfJ} is not needed, as
the definition of $K$ already includes an integral over angles.}{the
angular distribution is irrelevant and therefore there is no additional factor of
$4\pi$.} Both from the standpoint of developing intuition and computational convenience it is useful to explicitly
expand $K^*$ in the kink number. We introduce the WLC closed spatial propagator convolutions which we denote
\begin{equation}
\pnlabel{kn}
{\cal K}^{(\kn )} \equiv \left[ K\otimes\right]^\kn (0;L),
\end{equation}
where again  the $\otimes$ represents both spatial and arc length
convolutions. The computation of the ${\cal K}^{(\kn )}$ is discussed
in section \ref{kncomp}. In terms of the ${\cal K}^{(\kn )}$, the
free tangent $J$ factor is
\begin{equation}
\pnlabel{ftjf}
J_L^* = e^{-\zeta L} \sum_{\kn =0}^{\infty} \zeta^\kn {\cal K}^{(\kn +1)},
\end{equation}
where we have defined the  ${\cal K}^{(\kn )}$ to be independent of the
kinking parameter $\zeta$.
The kink number sum is illustrated with a diagrammatic expansion in fig \ref{kink.fig}.
The probability of the $\kn $ kink state can be concisely written in terms of the ${\cal K}^{(\kn )}$
\begin{equation}
\pnlabel{ftjpfd}
{\cal P}_\kn  = e^{-\zeta L}\zeta^\kn \frac{ {\cal K}^{(\kn +1)}}{J_L^*}.
\end{equation}
This expression can be interpreted as the kink number distribution
for a looped chain, a constraint that induces kinking in a manner
roughly analogous to the tangent constraints already discussed in
detail. The looping $J$ factor is plotted in fig \ref{jp}. In this
figure, we can see that the intuition we developed computing the
moment-bend constitutive relation is borne out in the looping $J$
factor, despite the fact that the process is thermally driven. In
the short-length limit, the ability of the chain to kink
dramatically reduces the bending energy and increases the looping
$J$ factor. In the short-length limit, a single kink is nucleated
in a manner almost exactly analogous to the process we have
described in detail for the moment-bend constitutive relation. We
will discuss these results and their scaling in more detail after
computing the KWLC $J$ factor.

\subsection{The cyclization $J$ factor}
\begin{figure}
\includegraphics{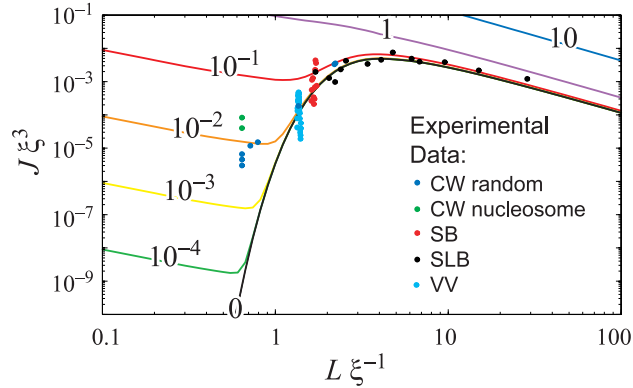}
\caption{\pnlabel{widomdata} (Color.) The KWLC cyclization $J$ factor as a
function of contour length $L$ for various values of the kink
parameter $\zeta$. As
discussed in the text, our theory does not include the twist
induced 10.5 bp modulation of the $J$ factor. The numbers labeling
the curves indicate the value of the dimensionless parameter
$\xi \zeta$. The WLC theory corresponds to $\zeta=0$. For
large contour length $L$, the effect of kinking can be accounted
for by computing the $J$ factor using the effective persistence
length, $\xi \zeta$. As the contour length $L$ falls below a
persistence length, kinking dramatically increases the $J$ factor,
even for small $\zeta$. For small $L$ the chain is two kink
dominated and diverges, in contrast to the WLC theory which
precipitously falls to zero at small $L$. Experimental cyclization
data for DNA are plotted for comparison, assuming $\xi=50$ nm.
(Data sources: CW\cite{CW}, SB\cite{Shore83a}, SLB\cite{Shore81},
and VV\cite{Vologodskaia2002}.) At contour length $L=0.6\xi$, the
experimentally measured $J$ factor is $\approx10^4$ times larger
than predicted by the WLC theoretical curve. The KWLC with
$\zeta\xi=10^{-2}$ correctly captures this behavior, while matching the WLC
theory at large contour length. }
\end{figure}
Although the computation of the free tangent $J$ factor is more
direct and intuitive, the $J$ factor with tangent alignment is of
more phenomenological interest. The computation begins with the
tangent-spatial partition function defined in eqn \ref{tsp} for
end distance zero and aligned tangents. Since the transformed WLC
tangent-spatial propagator is unknown, it would initially appear
to preclude exploiting the exact results derived above. But
intuitively we know the chain may be closed at any point resulting
in an identical $J$ factor. For kinked chains, it is convenient to
close the chain at a kink where the tangent alignment condition is
no longer required. The only chains for which this simplification
cannot be applied are the unkinked chains for which the $J$ factor
is already known\cite{SY}. To show this explicitly, we write the
transform of Eqn. \ref{tsp} in its expanded form and specialize to
$\vec{t}=\vec{t}_0$ to obtain the cyclization partition function:
\begin{equation}\pnlabel{e:cpf}
{\cal Z}^*_C(L)=G(0;\vec{t},\vec{t};L)+\sum_{\kn =0}^\infty\left( \zeta
G'(\vt\ )\otimes\left(\zeta
K\otimes\right)^\kn G'(\vt\ )\right)(0;L).
\end{equation}
We can now use the composition property of the propagators to
replace the initial and final tangent-spatial propagators with a
single spatial propagator. Some care is now required in performing
the convolution, as described in  appendix~\ref{a:fltc1}. The
cyclization partition
function then becomes (Eqn. \ref{circ.conv.eqn})
\begin{equation}
\pnlabel{ccpf}
{\cal Z}_C^*(L) = G(0;\vec{t},\vec{t};L) + \frac{1}{4 \pi}
\sum_{\kn =1}^{\infty}\frac{\zeta^\kn L}{\kn }{\cal K}^{(\kn )},
\end{equation}
where we have expressed the result in terms of the zero end
distance spatial propagator convolutions, ${\cal K}^{(\kn )}$. The
kink number sum is illustrated with a diagrammatic expansion in
fig \ref{kink.fig}. This equation has an analogous form to the
looping $J$ factor in Eqn. \ref{ftjf}. The only complication here
is that for kinked chains, the state counting has subtly changed
since we close the chain at a kink.  For inverse transform
numerical computations, it is convenient to write a transformed
partition function
\begin{equation}
\pnlabel{fltcpf}
\tilde {\cal Z}_C^*(k;p) = \tilde  G+\frac{1}{4\pi}\frac{\partial}{\partial p}\log \left[1-\zeta \tilde K\right],
\end{equation}
although the expression is understood to only have physical meaning when the chain is closed ($\vec{x}=0$).
Our derivation of the cyclized partition function implies that the
KWLC tangent-spatial propagator is known for one special case
\begin{equation}
G^*(0;\vec{t},\vec{t};L) = e^{-\zeta L} {\cal Z}_C^*(L),
\end{equation}
which is precisely the expression we need to compute the $J$ factor.
In terms of the KWLC tangent-spatial propagator, the KWLC $J$ factor is
\begin{equation}
\pnlabel{jexp2}
J^* = 4\pi G^*(0;\vec{t},\vec{t};L) = e^{-\zeta L}\sum^{\infty}_{\kn =0}
\zeta^\kn  {\cal J}^{(\kn )},
\end{equation}
where we have explicitly expanded the $J$ factor in kink number.
The ${\cal J}^{(\kn )}$ are defined by
\begin{equation}
\pnlabel{jfexp}
{\cal J}^{(\kn )} \equiv \begin{cases} J, & \kn =0 \\
                                    L\kn ^{-1}{\cal K}^{(\kn )}, & \kn >0 \end{cases}.
\end{equation}

\Fref{widomdata} compares experimental data to our theoretical
calculation of $J^*$. Details of the calculation are discussed in appendix~\ref{kncomp}.
\pn{Note that setting the kink
density to $\zeta\approx10^{-2}/\xi=0.2/\mu$m roughly reproduces the
experimental cyclization data. \Eref{zeta}
connects $\zeta$ to the density of vertices $\ell$ and the free
energy cost $\epsilon$ of creating a kink. Assuming that the site density is
just the DNA base pair length $\ell=0.34\,$nm, we can estimate the kink energy,
\begin{eqnarray}\label{eke}
\epsilon =  \ln \left[\frac{2\xi}{\ell^2\zeta}\right] \approx 15\
kT\approx9\,\mbox{kcal/mol}\ .
\end{eqnarray}
Although we do not discuss detailed microscopic models in this paper,
it is interesting to note that
molecular modeling studies have found that in B-form DNA, base pairs
indeed open individually and noncooperatively with an activation energy of 10--20kcal/mol~\cite{bern97a}.}{}

\subsection{Topologically induced kinking}
It is useful to discuss kink number in chains that are
topologically confined to be cyclized. These chains have both the
kink inducing tangent and spatial constraint.
\begin{figure}
\includegraphics{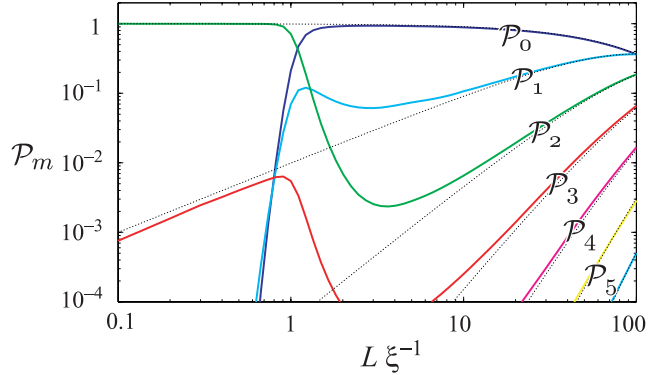}
\caption{\pnlabel{pndist}
The kink-number distribution compared for cyclized chains (solid curves) and
unconstrained chains (dashed curves) as a function of contour length $L$.
To illustrate constraint-driven kinking, we have chosen the
illustrative value
$\zeta\xi=10^{-2}$. At large contour length $L$, the cyclization constraint
is irrelevant but as the arc length shrinks to roughly a persistence length,
the bending energy required to cyclize the chain becomes significant and
there is a dramatic transition to the two kink state which dominates at short
contour length. The contributions of one and $\kn >2$ kink states are secondary.
}
\end{figure}
We can write the kink number distribution concisely in terms of
the $J$ factor
\begin{eqnarray}
{\cal P}_\kn  =  e^{-\zeta L}\zeta^\kn  \frac{{\cal J}^{(\kn )}}{J^*},
\end{eqnarray}
which is analogous to eqn \ref{ftjpfd} and depicted in fig \ref{pndist}.

The effects of kinking on the $J$ factor are dramatic even when
the kinking parameter $\zeta$ is small!  \Fref{widomdata} shows that the
WLC $J$ factor precipitously decreases with loop contour length
due to the increasing elastic energy required to close the loop:
\begin{equation}
J \sim e^{-2\pi^2\xi/L}.
\end{equation}
In dramatic contrast, the KWLC $J$ factor, while tracking the WLC $J$ factor at large contour
length, turns over at small contour length and increases divergently.
Physically, this small contour length divergence can be understood as an increase in the ratio of available
cyclized to noncyclized states which is roughly inversely proportional to the physical volume explored by the
chain when one end is fixed.
This divergent increase in the density of states can also be seen for the Gaussian Chain
in the short length limit, although this limit is not physical for polymer systems. For the Gaussian chain,
the $J$ factor is monotonically decreasing with contour length since the only obstruction to the
ends finding each other is entropic, and the number of available noncyclized states scales like $L^{3/2}$
(eqn \ref{gcjf}).
For KWLC, once the chain is kinked twice, it is advantageous to shorten the chain which, while decreasing the
degeneracy of the first kink location, $\propto L$ (eqn \ref{jfexp}), increases the density of cyclized states,
$\propto L^{-2}$
(eqn \ref{k2rr}).
Therefore, there is a net $L^{-1}$ scaling of the two kink term ${\cal J}^{(2)}$ (eqn \ref{jfexp}).
In this limit, the contributions from chains with kink numbers greater than two scale like $L^{\kn -3}$
(eqns \ref{jfexp} and \ref{knrr}),
implying that at short lengths the two kink term dominates.
In addition, in most physically interesting scenarios the kinking
parameter $\zeta$ is small. The probability of $\kn $ kink number state scales roughly like the average kink number
for the unconstrained chain to the $\kn $th power $(L\zeta)^\kn $ (eqn \ref{jexp2}), which further decreases the importance of
higher kink number states. The dramatic transition from the zero kink to the two kink state at short
contour length is evident in fig \ref{pndist}. Interestingly, recent
molecular-dynamics simulations on a 94~bp DNA minicircle indicate the presence
of  two sharply kinked regions~\cite{laveryPrivate}.

Physically, we can understand the onset of this transition by
roughly comparing the free energies of the two kink term and the zero kink term to find the length at
which these two are roughly equal. Here we merely wish to motivate our results as clearly and simply as
possible so we shall ignore the difference in the density of states, even though its effect is
quantitatively important. We therefore treat the free energy of the zero kink term as the bending energy
only and the free energy of the two kink term as twice the kink energy
in the discrete model (eqn \ref{zeta}):
\begin{equation}
2\epsilon  \sim \frac{2\pi^2\xi}{L_{\rm crit}}.
\end{equation}
When the bending energy equals the energy required to nucleate two
kinks, the transition occurs. It is important to remember that the
$\zeta$-dependence is relatively weak while the bending
energy scales like the inverse of the contour length. Below $L
\sim \xi$, the bending energy grows divergently implying that even
for very small kink densities, kinking always becomes important at
short enough contour length. That is to say, we are almost assured
of observing elastic breakdown effects below contour lengths of
roughly a persistence length.

Previously we had shown that for moment-bending, only a single
kink was nucleated, in contrast to the current example where the
one kink contribution to the $J$ factor, ${\cal J}^{(1)}$, is of
little scaling importance since the chain must still bend as
illustrated schematically in fig \ref{kink.fig}. This will not be
the case for the KWLC looping $J$ factor, $J_L^*$, which lacks the
tangent constraint implying that only one kink is required to
relieve all the elastic bending. The one kink term of $J_L^*$
therefore diverges like $L^{-2}$ (eqn \ref{k2rr}) as explained
above. $J_L^*$ is plotted in fig \ref{jp}.

\section{Discussion}
\pnlabel{discussion}Our main results are summarized in \frefs{force.ext}, \ref{philfig},
\ref{widomdata}, and \ref{pndist}. We formulated a generalization of the WLC model, in
which a semiflexible polymer can develop flexible sites of an
alternative conformation. We found that taking the density of kinks
in the unstressed polymer to be about $0.01$ per persistence length
has negligible effect on the force--extension relation, but vastly
enhances the probability of cyclization for chains shorter than
a persistence length, as seen in recent
experiments on DNA~\cite{CW}.

Various microscopic mechanisms could furnish the kinking mechanism in
DNA, for example single-basepair flipout or strand separation
\cite{wiggins2}. But any complete, microscopic analysis of high-curvature DNA
conformations would also have to include a variety of effects, for
example those arising
from the significant thickness of the DNA molecule on the few-nanometer scale
of a short circle, strong polyelectrolyte effects, and so on. We have
taken the attitude that any net nonlinear softening at high curvature
will lead to generic new phenomena.
By summarizing all such effects into a
single phenomenological parameter, our model focuses attention on the
general mesoscale physics of kinking.  The KWLC's
generality also makes it a
useful starting point for studying the conformations of other stiff
biopolymers, such as actin.

Other diagnostics of low-curvature physics, for example light
scattering, also turned out to be almost indistinguishable from the
linearly-elastic Wormlike Chain model.  It is only by conducting
experiments that are explicitly sensitive to high curvature, that
we can measure the nonlinear response to bending. DNA cyclization
offers one
experimentally tractable measurement sensitive to the high-curvature
physics of free DNA in solution; we gave predictions
for other, future tests, for example the moment--bend relation and
the kink number as a function of constraints.
Indeed, the predicted average number of kinks has direct structural
implications for very small DNA loops, and for processes involving
such loops, for example, the looping implicated in some
gene-regulatory
mechanisms~\cite{bell88a,Rippe1995,Muller1,Muller2,Rippe2001,Zhang}.
Recent simulations indeed suggest that spontaneous kinking may play a
role in such situations~\cite{laveryPrivate}.

Some mathematical aspects of the model, for example
the divergence of the theoretical $J$ factor at small
contour length (\fref{widomdata}), are artifacts of the simplified
picture we have proposed. The small contour length
divergence is due to the two kink term, which can close a loop without
elastic bending regardless of the contour length, via the creation
of 180-degree kinks! Clearly due to the finite thickness of the
chain, this divergence is unphysical. We can consider a number of
modifications to the theory to fix this problem: kink angle cutoffs, kinks that are not perfectly flexible, etc.
But all these proposals require adding additional parameters to the
model, rendering it both less
tractable and less predictive,
since the additional parameter must then be fit to experimental data.

The KWLC is  in
essence a coarse-grained, effective theory for systems where
kinking occurs and the kinks are localized compared with the chain
persistence length. Its virtue is that it offers a simple way to
characterize stiff biopolymers, and a quantitative guide to the mesoscale effects of kinking.
Thanks to this simplicity,
we were able to compute many
results in this paper exactly, without extensive numerical simulation.
We discuss the specific application of
KWLC to DNA at length elsewhere\cite{wiggins2}.

\begin{acknowledgments}
We thank J. Widom and T. Cloutier for sharing their data and
insights on DNA bending before they were generally available. PAW
thanks A. Spakowitz for insightful conversations and his results
before they were generally available, and T.-M. Yan for his
careful instruction in the mysteries of quantum mechanics long
ago. The authors would also like to acknowledge M. Inamdar, W.
Klug, J. Maddocks and Z.-G. Wang for helpful discussions and
correspondence, and Yongli Zhang for comments on the draft
manuscript. We acknowledge grant support from an NSF graduate
fellowship [PAW]; the Human Frontier Science Foundation and NSF
grant DMR-0404674 [PN]; the Keck Foundation and NSF Grant
CMS-0301657 as well as the NSF-funded Center for Integrative
Multiscale Modeling and Simulation [RP and PAW].
\end{acknowledgments}


\appendix

\section{Fourier \& Laplace Transforms \& convolution
theorems\label{a:a}}
The relations listed below are well known\cite{arfken} but essential to our derivations.
We define the 3D spatial Fourier Transform and inverse transform
\begin{eqnarray}
\tilde F(\vec{k}) &\equiv& {\cal F}\{ F(\vec{x}) \} = \int d^3x\ F(\vec{x}) e^{-i\vec{k}\cdot\vec{x}}, \\
F(\vec{x}) &=&  {\cal F}^{-1}\{ \tilde F(\vec{k}) \} = \left(\frac{1}{2\pi}\right)^3\int d^3k\ \tilde F(\vec{k}) e^{i\vec{k}\cdot\vec{x}}.
\end{eqnarray}
The Faltung theorem states that Fourier Transform of a convolution is the product of the Fourier Transforms:
\begin{equation}
{\cal F} \{F\otimes G\} = \tilde F \tilde G,
\end{equation}
for functions $F$ and $G$ where the spatial convolution is defined
\begin{equation}
F\otimes G(\vec{x}) \equiv \int d^3x'\ F(\vec{x}{\,}')G(\vec{x}-\vec{x}{\,}').
\end{equation}
The generalization of the Faltung theorem is true for $n$ functions
\begin{equation}
\pnlabel{ftft}
{\cal F} \{F_1\otimes ... \otimes F_m\} = \tilde F_1 ... \tilde F_m.
\end{equation}

We define the 1D contour length Laplace Transform and inverse
transform
\begin{eqnarray}
\pnlabel{laplace}
\tilde F(p) &\equiv& {\cal L}\{ F(L) \} = \int^{\infty}_{0}dL\ F(L) e^{-pL}, \\
F(L) &=&  {\cal L}^{-1}\{ \tilde F(p) \} = \frac{1}{2\pi i}\int_{\cal L}dp\ \tilde F(p) e^{pL},
\end{eqnarray}
where $\int_{\cal L}$ denotes a contour integral along the Laplace contour.
The Faltung theorem states that Laplace Transform of a convolution is the product of the Laplace Transforms:
\begin{equation}
{\cal L} \{F\otimes G\} = \tilde F \tilde G,
\end{equation}
for functions $F$ and $G$, where the arc length convolution is defined
\begin{equation}
F\otimes G(L) \equiv \int_0^LdL'\ F(L')G(L-L').
\end{equation}
The generalization of the Faltung theorem is true for $m$ functions
\begin{equation}
\pnlabel{ftlt}
{\cal L} \{F_1\otimes ... \otimes F_m\} = \tilde F_1 ... \tilde F_m\ .
\end{equation}

\subsection{Circular convolutions\pnlabel{a:fltc1}}
\begin{figure}
\includegraphics{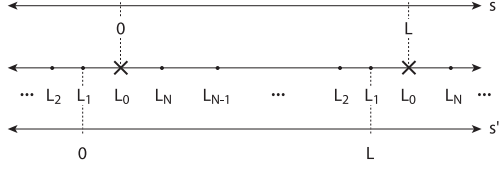}
\caption{\pnlabel{circ.conv} A schematic diagram of the coordinate transformation exploited to
compute the circular convolution. The crosses represent chain ends and dots represent kinks.
The center line represents a periodic coordinate system. For regular convolutions, we set
the chain end to be the zero and we compute the convolution in the $s$ coordinate system.
For the circular convolution, it is more convenient to choose $L_1$, the first kink arc length
position as the zero and sum over the chain end positions as represented by the $s'$ coordinate system.}
\end{figure}
For the closed chain we need to evaluate a special type of convolution which is circular.
By circular we mean that the end points are identified so that the arc-length
position of the chain ends is not at an end point of the propagator. In this case it is
convenient to redefine the arc-length coordinate system to be zero at the position of the first kink,
$L_1$, and sum over the position of the chain ends as depicted in fig
\ref{circ.conv}. The $\kn$ kink
contribution to the partition function is therefore
\begin{equation}
{\cal Z}_\kn^* = \frac{\zeta^\kn
}{4\pi}\left[\left(K\otimes\right)^{\kn-1}LK\right](0;L),
\end{equation}
where the factor of $L$ comes from the integration over the end position and varies in the convolution.
To simplify this result it is convenient to go to the transformed partition function
\begin{eqnarray}
\tilde {\cal Z}_\kn^* = -\frac{\zeta^\kn }{4\pi} \tilde
K^{\kn-1}\frac{\partial}{\partial p} \tilde K = -\frac{\zeta^\kn }{4\pi \kn} \frac{\partial}{\partial p} \tilde K^\kn ,
\end{eqnarray}
where the $L$ has been transformed into a $p$ derivative. We now return from Fourier-Laplace space, giving:
\begin{equation}
\pnlabel{circ.conv.eqn}
{\cal Z}_\kn^* = \frac{\zeta^\kn L}{4\pi \kn}\left[K\otimes\right]^{\kn}(0;L),
\end{equation}
 which is written in terms of convolutions of the spatial propagator only.

\subsection{Numerical inverse transforms}
\pnlabel{nit} To compute the numerical inversions of the Fourier
and Laplace Transforms involving the spatial propagator, we first
truncate the continued fraction \cite{Andy} in \eref{kprop}, then
we compute the numerical inversion with the built-in Mathematica
functions {\tt InverseLaplaceTransform} and {\tt
InverseFourierTransform}. In particular, the structure factor and
partition function in an external field involve only a single
numerical inverse Laplace Transform.

\subsection{$J$ factor computation}
\pnlabel{kncomp} We have chosen to present most of our results in
the last section as explicit series in kink number rather than
writing them in the summed form (eqn \ref{fltsp} and
\ref{fltcpf}). The purpose is two fold. First these expansions
allow $J$ to be computed efficiently for many small values of
$\zeta$, because the ${\cal K}^{(\kn )}$ are independent of $\zeta$ and
only the first few must be computed explicitly for sufficient
numerical accuracy. Furthermore the ${\cal K}^{(\kn )}$ are simply
related to the kink number distribution allowing the same
computation to suffice for both results. Our computational
discussion will mainly focus on the short contour length limit
where these kink number expansions converge quickly. As we have
already discussed at length, the large contour length limit can
trivially be computed with the WLC results using the renormalized
effective persistence length, $ \xst$. This corresponds to
the $k\rightarrow 0$ and $p\rightarrow 0$ limit where the theories
are identical. In fact, in this limit, we can use the Gaussian
chain to compute the $J$ factor. This computation appears briefly
in appendix \ref{gc}.

It is at short contour length where the two theories significantly diverge and kinking is induced.
As we have discussed above, in the limit as the contour length goes to 0, the polymer resembles a
rigid rod. It is problematic to directly compute the inverse transforms of $K^*$ or $K$
numerically in this limit since
\begin{equation}
K(0;L\rightarrow 0) \propto \int^\infty_0 dk\ k \sin kL,
\end{equation}
which, although it can be computed analytically by expanding sine
into two exponentials then integrating them on different contours,
is problematic numerically. It is therefore convenient to use the
expanded definition of $K^*$ in terms of the ${\cal K}^{(\kn )}$ (eqn
\ref{kn}). In the short contour length limit, an asymptotic
expression already exists\cite{YamakawaBook} for $J$ and
$K(0;L)$.

Convolutions of $K$ must still be computed. Numerical computations of the inverse transforms of
powers of $\tilde K$ are still problematic in the $L \rightarrow 0$ limit since,
while they are more convergent than $K$, they must be integrated to large $k$ where it is
difficult to compute accurate Laplace transforms. But this implies it is in precisely this
limit that bending is really irrelevant and that the mechanics of these kinked chains
becomes kink dominated. Once there are two or more kinks, the chain can now be closed without
elastic bending. That is to say that, when these terms become difficult to calculate numerically,
they can be well approximated by the kinkable rigid rod! The kinkable rigid rod is treated in
appendix \ref{krr}. As a practical matter there is a small contour length regime ($L \sim 1$),
between the rigid rod limit and the contour length at which numerical transform inversion
are rapidly convergent and where it is most convenient to use direct Monte Carlo integrations
to compute the ${\cal K}^{(\kn )}$. These direct Monte Carlo integrations serve as a useful check
on our other numerical and analytic computations.
For large dimensionless kink densities, the kink-number expansion is not rapidly converging and
direct numerical inversions of the exactly summed transformed results are required.
For most computations of the $J$ factor at small kink density, the rigid rod approximation suffices to compute the
two kink term and the kink number sum can be truncated at this point as illustrated schematically by \Fref{pndist}.

\section{Kinkable rigid rod}
\pnlabel{krr}
In this section we develop the theory of kinkable rigid rods, the infinite persistence length limit of the KWLC.
This theory is useful for discussing the short loop limit of the $J$ factor.
The rigid rod tangent-spatial propagator is
\begin{equation}
G(\vec{x},\vec{t}_L,\vec{t}_0,L) = \delta^{(3)}\left[\vec{x}-L\vec{t}_L\right]\ \delta^{(2)}\left[\vec{t}_L-\vec{t}_0\right].
\end{equation}
The spatial propagator, $K$, is obtained by averaging and summing over the two tangents (eqn \ref{spe})
\begin{eqnarray}
K(\vec{x},L) &=& \frac{1}{4\pi L^2} \delta\left[|\vec{x}|-L\right].
\end{eqnarray}
The Fourier and Fourier-Laplace transform spatial propagator are
\begin{eqnarray}
\tilde K(\vec{k},L) &=& \frac{\sin kL}{kL}, \\
\tilde K(\vec{k},p) &=& \frac{1}{k} \tan^{-1} \frac{k}{p}.
\end{eqnarray}
In order to discuss the $J$ factor limit, we will need the convolutions
\begin{equation}
{\cal K}^{(\kn)} \equiv [K\otimes]^\kn (0,L),
\end{equation}
which is proportional to the probability density of the end distance be $0$ after arc length $L$ and $\kn -1$ kinks.
${\cal K}^{(1)}$ is zero since the rod is rigid and the ends cannot meet unless the chain kinks.

${\cal K}^{(2)}$ is also fairly straight forward. It is convenient to compute the convolution explicitly
\begin{eqnarray}
{\cal K}^{(2)} &=& \frac{1}{2\pi^2} \int^L_0 dL_1\int^\infty_0dk\ k^2\ \frac{1}{kL_1}\sin kL_1 \ \frac{1}{k(L-L_1)}\sin k(L-L_1)  \\
\pnlabel{k2rr}
&=& \frac{1}{2\pi L^2},
\end{eqnarray}
where the Fourier transform delta function has been used to evaluate the integral.

The computation of ${\cal K}^{(3)}$ requires some care. Again it is convenient to compute the convolution explicitly
\begin{eqnarray}
\nonumber
{\cal K}^{(3)} &=& \frac{1}{2\pi^2} \int^L_0 dL_1\int^L_0 dL_2\int^L_0 dL_3\ \int^\infty_0dk\ k^2\ \times \\
& & \frac{1}{kL_1}\sin kL_1 \ \frac{1}{kL_2}\sin kL_2\  \frac{1}{kL_3}\sin kL_3 \ \delta \left( L_1+L_2+L_3-L\right), \\
&=& \frac{1}{8 \pi^3} \int^{L/2}_0 dL_1\int^{L/2}_{L/2-L_1} dL_2 \frac{1}{(L-L_1-L_2)L_1L_2}, \\
&=& \frac{\pi}{16L}.
\end{eqnarray}

For convolution number $\kn >3$, we exploit the Fourier-Laplace transform method
\begin{eqnarray}
{\cal K}^{(\kn )} &=& \frac{1}{2\pi^2} \int^\infty_0dk\ k^2\int_{\cal L} dp\ \left(\frac{1}{k} \tan^{-1} \frac{k}{p} \right)^\kn  e^{pL}\\
 &=& \frac{1}{2\pi^2} \int^\infty_0dk'\ {k'}^2\int_{\cal L} dp\ p^{3-\kn }\left(\frac{1}{k'} \tan^{-1} k' \right)^\kn  e^{pL},
\end{eqnarray}
where we have made the substitution $k'=k/p$. Now let us compute
the $k'$ integral, which must be done numerically. We now make the
substitution $\tan x = k'$. The integral in $k'$ becomes
\begin{equation} I_\kn  \equiv \frac{\kn }{\kn -3} \int^{\pi/2}_0dx
\tan^{3-\kn } x \ x^{\kn -1},
\end{equation}
which we computed using \textsl{Mathematica}.
\begin{table}
\begin{tabular}{|c|c|}
\hline
convolution     & Integral  \\
number $\kn $      & $I_\kn $ \\
\hline
4 & 2.249 \\
5 & 0.841 \\
6 & 0.461 \\
7 & 0.300 \\
$\cdots$ &$\cdots$ \\
\hline
\end{tabular}
\caption{\pnlabel{table}   Values for the numerically computed integral $I_\kn $ for the first few $\kn $.}
\end{table}
The $p$ integral is now a simple contour integral which gives
\begin{eqnarray}
\pnlabel{knrr}
{\cal K}^{(\kn )} &=& \frac{I_\kn }{2\pi^2} \frac{L^{\kn -4}}{(\kn -3)!},
\end{eqnarray}
for $\kn >3$. The first few values of $I$ are computed numerically in table \ref{table}.

The kinkable rigid rod theory, derived above, provides a very useful analytic check of the
KWLC model at short contour length. For short cyclized polymers, the bending between kinks can be
ignored since these segments are significantly shorter than a persistence length. As we have
illustrated above, the computation of the dominant two kink contribution is straightforward
in this limit.

\section{Gaussian limit}
\pnlabel{gc}
The Gaussian limit provides a useful analytic limit to the KWLC theory for long contour length.
In this limit, the length of the polymer makes the initial tangent condition irrelevant and describes the
spatial distribution for chain extensions short compared with the contour length.

For large $L$, we can work with the Gaussian distribution.
The Gaussian distribution is
\begin{equation}
G(\vec{x};\vec{t},\vec{t}{\ }';L) = \frac{1}{4\pi}\left(\frac{3}{4\pi \xi
L}\right)^{3/2} \exp\left[-\frac{3\vec{x}^{\, 2}}{4\xi L}\right]
\end{equation}
for persistence length $\xi$. The $J$ factor is
\begin{equation}
\pnlabel{gcjf}
J = 4\pi G(0;\vec{t},\vec{t};L)= \left(\frac{3}{4\pi \xi L}\right)^{3/2}.
\end{equation}
For KWLC, the persistence length is replaced by the effective persistence
length $ \xst$:
\begin{equation}
J^* = 4\pi G^*(0;\vec{t},\vec{t};L)= \left(\frac{3}{4\pi { \xst} L}\right)^{3/2}.
\end{equation}
In the Gaussian limit, the convolution functions ${\cal K}^{(\kn )}$ can be computed without difficulty:
\begin{equation}
{\cal K}^{(\kn )} = \frac{L^{\kn -1}}{(\kn -1)!}\left(\frac{3}{4\pi \xi L}\right)^{3/2}.
\end{equation}
But this expression holds only when the number of kinks is small.

\section{Summary of notation}\pnlabel{notation}
We imagine a chain of total contour length $L$, with $L/\ell$ elementary segments of
length $\ell$. Individual segments will be referred to by their sequence
number $n=0,\ldots N$, where $N=(L/\ell)-1$, or by arclength $s=n\ell$.
A configuration $\Gamma$ consists of a sequence of tangent vectors
$\{\vt_0,\ldots,\vt_N\}$.

The stiffness parameter (WLC persistence length) $\xi$, one-vertex partition function $\pnQ$, kink formation energy
$\epsilon$, and kinking parameter $\zeta$ are defined in
\srefs{model}--\ref{partitionfunction}.  $\xst$ and
$\pnQ^*$ are related quantities relevant to the KWLC. The kink length
is $\xk  \equiv \zeta^{-1}$.

The measure $d^2\vt\,$ denotes solid angle on the sphere of unit
vectors $\vt$. Square brackets denote the functional measure
$\pnmeas{\vti}$; see \eref{normalization}.

The partition functions ${\cal Z}(L)$ and ${\cal Z}(\vtf,\vti;L)$ refer
to unconstrained and constrained functional integrals over a chain of
length $L$ in the continuum limit. Rotation invariance implies that
the constrained function depends only on the angle $\theta$ between
the vectors, so we sometimes write it as ${\cal Z}(\theta;L)$.
Discretized versions of the partition functions are denoted
with the subscript ``discrete,''  and KWLC versions with a star.
Related quantities include the free energy $F(\theta;L)=-\log{\cal
Z}(\theta;L)$ (\eref{e:dfF}) and the
normalized tangent partition function
(or propagator) $H(\vtf,\vti;L)={\cal Z}(\vtf,\vti;L)/{\cal Z}(L)$.
Laplace transforms of these functions on $L$ are denoted with a tilde.

When it is important to maintain spatial information, we introduce
space-dependent functions ${\cal Z}(\vx,\vtf,\vti;L)$
(\eref{kwlcpfd}), $K(\vec{x};L)$ (\eref{spe}), and
$G'(\vec{x},\vec{t};L)$ (\eref{e:gprime}). Fourier--Laplace transforms
of these functions on $\vx,L$ are again denoted with a tilde.

Laplace and Fourier transformations, and the corresponding
convolution operation $\otimes$, are defined in appendix~\ref{a:a}.
Repeated convolutions of $K$ give the functions ${\cal K}^{(\kn)}$ (\eref{kn}),
and the related ${\cal J}^{(\kn)}$ (\eref{jfexp}).

The partition function in an external force is ${\cal Z}_{\vec{f}}$
(\eref{fepf}); the
cyclization partition function is ${\cal Z}^*_C(L)$ (\eref{e:cpf}).

In an expansion in kink number, $\kn$ labels the number of kinks and
$i=1,\ldots,\kn$ labels which kink is in question. The kinks are taken
to be located at $n_{i}$, or at arc length position $L_i=\ell n_i$.

\end{document}